

\input phyzzx.tex

\tolerance=10000
\def\pslash{{p\hskip -4.5pt{/}}}
\def\psh{{\rlap{/}{p}}}
\def\msh{{\rlap{/}{m}}}
\def\nsh{{\rlap{/}{n}}}
\def\Dslash{{D\hskip -7pt{/}}}
\def\lmdash{{\Lambda\llap{/}}}
\def\Sigmash{{\Sigma\llap{/}}}
\def\np{Nucl. Phys.}
\def\pl{Phys. Lett.}
\def\pr{Phys. Rev.}

\def\ijmp{Intern. J. Mod. Phys.}
\def\mpl{Mod. Phys. Lett.}
\def\nc{Nuovo Cim.}

\def\cqg {Class. Quantum Grav.}
\def\jmp {J. Math. Phys.}
\def\sovjnp {Sov. J. Nucl. Phys.}

%

\def\half {\textstyle {1 \over 2 }}
\def\seventh {\textstyle {1 \over 7 }}

\def\inttau {\int\! d\tau}
\def\xmu {x^\mu}
\def\xplus {x^{+}}
\def\amu {A_\mu}
\def\amin {A_{-}}
\def\aplus {A_{+}}
\def\aplmi {A_{\pm}}

\def\anut {A_0}
\def\anine {A_9}

\def\Phia {\Phi^A}
\def\phiA {\phi^A}
\def\phiB {\phi^B}
\def\hatphiA {{\hat\phi}_A}

\def\thetaA {\theta_A}
\def\thetaB {\theta_B}
\def\thetaC {\theta_C}

\def\PsiA {\Psi_A}
\def\PsiB {\Psi_B}
\def\PsiC {\Psi_C}
\def\psiA {\psi^A}
\def\Psimu {\Psi_\mu}

\def\psia {\Psi^a}
\def\Psia {\Psi_a}
\def\Psiplus {\Psi_{+}}
\def\Psimin {\Psi_{-}}
\def\psiadot {\Psi^{\dot a}}
\def\adot {\dot a}
\def\bdot {\dot b}
\def\grad {\partial}
\def\pmin {p^{-}}
\def\pplus {p^{+}}
\def\pmu {p^\mu}
\def\Psii {\Psi^i}
\def\Psij {\Psi^j}
\def\Psik {\Psi^k}

\def\gamplus {\gamma^{+}}
\def\gammu {\gamma^\mu}
\def\Gammu {\Gamma^\mu}

\def\gammai {\gamma^i}
\def\gammaj {\gamma^j}
\def\gammaij {\gamma^{ij}}
\def\gammaijk {\gamma^{ijk}}
\def\Lamnrs {\Lambda_{\mu\nu\rho\sigma}}
\def\Sigmnrs {\Sigma_{\mu\nu\rho\sigma}}

\def\Gamnrs {\Gamma^{\mu\nu\rho\sigma}}
\def\Gamsmn {{\Gamma^{\sigma\mu}}_\nu }
\def\Lamijk {\Lambda_{ijk}}
\def\rhomunu {\rho^{(\mu\nu )}}
\def\rhoij {\rho^{(ij) }}
\def\rhosignu {\rho_{\sigma\nu )}}
\def\hatphi {{\hat\phi}}
\def\phimu {\phi^\mu}
\def\phinu {\phi^\nu}
\def\kapa {\kappa^A}
\def\dA {d^A}
\def\dB {d^B}
\def\dC {d^C}
\def\Upsimupha {{\Upsilon^\mu}_A }
\def\Cmupha {{C_\mu}^A}
\def\omegmunu {\omega^{\mu\nu}}
\def\Sigmunu {\Sigma^{\mu\nu}}
\def\chimuA {{\chi^\mu}_A}
\def\chimuB {{\chi^\mu}_B}
\def\chinuA {{\chi^\nu}_A}
\def\chisigA {{\chi^\sigma}_A}
%
%
\Pubnum = {QMW-93-07}
\date = {July 1993}
 \pubtype={}
\titlepage
\title { PARTICLES, SUPERPARTICLES AND SUPER YANG--MILLS.}
\author {
 Christopher M. Hull}
 \andauthor{
 Jose-Luis V\'azquez-Bello}

\

\address {Physics Department, Queen Mary and Westfield College,\break
	     Mile End Road, London E1 4NS,\break
		 UNITED KINGDOM.}

\abstract
{This paper is concerned with theories describing spinning particles that
are formulated in terms of actions possessing either local world-line
supersymmetry or local fermionic {\it kappa}  invariance.  These classical
actions are obtained by adding a finite number of spinor or vector coordinates
to the usual space-time coordinates. Generalizing to superspace leads to
corresponding types of \lq spinning superparticle'  theories in which the
wave-functions are superfields in some (generally reducible) representation
of the Lorentz group.  A class of these spinning superparticle actions
possesses  the same spectrum as ten-dimensional supersymmetric
Yang--Mills theory, which it is shown can be
 formulated in terms of either vector or spinor superfields satisfying
supercovariant constraints.  The  models under consideration include some
that were known previously and some new ones. }

\endpage
\pagenumber=1

\chapter { Introduction.}

The quantum mechanics of  a  free superparticle  in ten-dimensional space-time
is of interest both because of its relationship to ten-dimensional Yang--Mills
theory and because it provides a description of the massless sector of the
superstring.  There are many  formulations of ten-dimensional superparticle
dynamics, all describing a particle evolving along a world-line in some
superspace and possessing a fermionic {\it kappa} symmetry  (this is the local
world-line symmetry for which the parameter
is a fermionic space-time spinor\REF\siege{W.
Siegel, \cqg\ {\bf 2} (1985) L95.}\REF\siegea{W. Siegel,
\np\ {\bf B263} (1985) 93.}\REF\siegela{W. Siegel,
\pl\ {\bf B205} (1988) 257.}\REF\siegelc{W. Siegel,
in {\it Unified String Theories}, M.B. Green
and   D. J. Gross eds.
(World Scientific, Singapore 1986).}[\siege-\siegelc ]).
The original
superparticle theory \REF\casal{R. Casalbuoni,
\pl\ {\bf B293} (1976) 49;\nc\ {\bf A33}
(1976) 389.}\REF\freunda{P. G. O. Freund, unpublished,
as quoted in A. Ferber, \np\ {\bf B132} (1978) 55.}\REF\BrSch{ L. Brink
and J. H. Schwarz, \pl\ {\bf B100}
(1981) 310.}[\casal-\BrSch ] was quantized in the light-cone gauge
and shown to yield the same spectrum as that of  ten-dimensional super
Yang--Mills theory. However, this model has defied covariant quantization
as there is no satisfactory covariant gauge choice for the kappa symmetry.
It is generally the case that a manifestly covariant formulation of a quantum
theory provides a more elegant and geometric description than one in which
manifest covariance is absent.  The difficulty in quantising the superparticle
in a covariant manner is related to the absence of a gauge field for the
kappa symmetry in these models which means that the only available gauge
conditions are ones involving the superspace coordinates $\xmu$ and $\thetaA$.
The gauge choice $\dot\theta =0$ was proposed in \REF\gauge{U. Lindstr\"om,
M. Ro\v cek, W. Siegel, P. van Nieuwenhuizen  and
	   A. E. van de Ven, \pl\ {\bf B224} (1989) 285.}\REF\gaugea{W.
Siegel, {\it Proc. of the 1989
Texas A + M String Workshop.}}[\gauge,\gaugea] but this
ran into difficulties \REF\fish{J. Fisch and Henneaux, Bruxells Preprint ULB
TH2-89-04.}\REF\fisha{F. Bastianelli, W. Delius and E. Laenen,
			    \pl\ {\bf B229} (1989) 223.}[\fish,\fisha ].
These stem from the fact that $\dot\theta =0$ does not constitute
a good gauge slice, in the sense that there are field configurations
which cannot be transformed to any configuration satisfying the gauge
condition by a gauge transformation
\REF\mike{M. B. Green and C. M. Hull, {\it The Covariant Quantization of the
Superparticle}, in the \lq Proceedings of the 1989 International Workshop on
Superstrings', edited by R. Arnowitt, M.J. Duff and C.N. Pope,
(World Scientific, Singapore, 1990).}\REF\mikea{M. B. Green and
C. M. Hull, \np\ {\bf B344} (1990) 115.}[\mike,\mikea].
This situation is similar to that for the Nambu-Goto string, for which
it is also difficult to find a covariant gauge choice.  For the Nambu-Goto
string the resolution is to introduce a world-sheet metric, in which case
covariant gauges can be defined by imposing conditions on the metric
instead of on the coordinates.  Similarly, for the superparticle, the
resolution is  to introduce a gauge field for the kappa symmetry, so that
covariant gauge choices can be defined by imposing conditions on the
gauge field [\siege].
\REF\kallrah{  R. E. Kallosh and M. A. Rahmanov,
\pl\ {\bf B209} (1988) 233.}\REF\nissim{E. Nissimov,
S. Pacheva and S. Solomon,
\np\ {\bf B296} (1988) 462;  \np\
{\bf B297} (1988) 349.}There are
several formulations which involve a gauge field $\psiA$
for the kappa symmetry. The simplest of these, due to Siegel [\siege ],
introduces $\psiA$ and a momentum conjugate to $\thetaA$.  However, this model
has a spectrum which is not a $N=1$, $D=10$ super Yang--Mills multiplet,
but is a twisted
$N=2$, $D=10$ supergravity multiplet with negative norm states [\mikea].
Thus, it is not equivalent to the earlier model [\casal,\BrSch]
and is not of direct physical
interest  since its spectrum has negative norm states.
Nevertheless, it serves as an interesting model example since it can be
covariantly gauge-fixed by choosing the gauge
$\psi =0$, $e=constant$ where $e$ is the world-line einbein
[\mikea,\kallrah,\nissim].
A full covariant treatment  requires the use of the Batalin--Vilkovisky (BV)
procedure
\REF\BV{I. A. Batalin and G. A. Vilkovisky, \pl\ {\bf B102} (1981) 27;
					    \pr\ {\bf D28} (1983) 2567.}[\BV],
 which in this case requires an infinite number of ghosts for ghosts;
 the complete quantum theory was derived in this way
 in [\mikea].

 A modification of this model was proposed in
\REF\sieggya{W. Siegel, \pl\ {\bf B203} (1988) 79.}\REF\siegeb{I. A.
Batalin, R. E. Kallosh and A. Van Proeyen,
	    {\it Symmetries of Superparticles and Superstring Actions},
	    in Quantum Gravity, eds. M. A. Markov, V. A. Berezin and
	    F. P. Frolov (World Scientific, 1988).}[\sieggya,\siegeb] and
analysed in
\REF\MikS{A. Mikovi\'c and W. Siegel,
\pl\ {\bf B209} (1988) 47.}\REF\KVanPT{R. E. Kallosh,
A. Van Proeyen and W. Troost,
		       \pl\ {\bf B212} (1988) 428.}[\MikS,\KVanPT ].
\REF\mikky{U. Lindstr\"om, M. Ro\v cek, W. Siegel, P. van Nieuwenhuizen and
	   A. E. van de Ven, \jmp\ {\bf 31}
(1990) 1761.}A full Batalin-Vilkovisky quantization
of this model was attempted
in [\mikky], but it seemed that there was no solution
to the Batalin-Vilkovisky master equation for this model, so that
there appeared to be an obstruction to the quantization of this model.
However, even if it turned out that this model was quantizable using
the Batalin-Vilkovisky approach (for example, if it turned out that
the correct
\REF\GH{M. B. Green and
C. M. Hull, \mpl\ {\bf A18} (1990) 1399.}ghost structure
was different from that proposed in [\mikky]),
it seems that this model does not  give the required
super Yang--Mills spectrum; indeed, it was shown in [\GH] that,
if one simply assumed that a BRST charge existed,
enough could be deduced about its structure to calculate the
full BRST cohomology of the theory for  cohomology classes of low
 ghost number, and
 it was found in [\GH] that a very large class of low ghost-number
 BRST cohomology
 classes were in fact trivial.

To get around these problems, it has been suggested that the model should
further be modified by adding extra coordinates to the $(\xmu,\thetaA )$
superspace
\REF\siegec{L. Brink, in {\it Physics and Mathematics of Strings},
	    eds. L. Brink, D. Friedan and A. M. Polyakov,
	    (World Scientific, 1990).}\REF\rocek{A. Mikovi\'c, M.
 Ro\v cek, W. Siegel, A. E. van de Ven,
	   P. van Nieuwenhuizen and J. P. Yamron, \pl\ {\bf B235} (1990)
106.}\REF\spaces{F. E$\beta$ler,
E. Laenen, W. Siegel and J. P. Yamron,
	      \pl\ {\bf B254} (1991) 411.}\REF\spacesa{F. E$\beta$ler,
M. Hatsuda, E. Laenen, W. Siegel, J. P.
Yamron, T. Kimura and A. Mikovi\'c, \np\ {\bf B364} (1991) 67.}[\GH-\spacesa].
There are now a number of such models which can be covariantly quantized
and which give the super Yang--Mills spectrum. These models fall into two
classes, those in which the classical superspace has an infinite number of
coordinates \REF\fishb{E. A. Bergshoeff and R. E. Kallosh, \np\ {\bf B333}
(1990) 605.}\REF\fishc{E. A. Bergshoeff and R. E. Kallosh, \pl\ {\bf B240}
(1990) 105.}\REF\spacesd{E. A. Bergshoeff, R. E. Kallosh and A. Van Proeyen,
\pl\ {\bf B251} (1990) 128.}\REF\spacese{R. E. Kallosh, \pl\ {\bf B251} (1990)
134.}\REF\spacesf{M. Huq,
 Int. J. Mod. Phys. {\bf A7} (1992) 4053.}\REF\bello{ J.
L. V\'azquez-Bello,
 \ijmp\ {\bf A19} (1992) 4583.}[\rocek-\bello] and those in which it has a
finite number
[\sieggya,\GH,\rocek].  This paper will be restricted to
formulations with a finite number of classical superspace coordinates.
\REF\twistor{D. P. Sorokin, V. I. Tkach and D. V. Volkov,   \mpl\ {\bf A4}
(1989) 901;  D. P. Sorokin, V. I. Tkach and D. V. Volkov and
A. A. Zheltukhin,  \pl\ {\bf B216} (1989) 302;
A. I. Gumenchuk and D. P. Sorokin, \sovjnp\ {\bf 51} (1990) 350;
F. Delduc and E. Sokatchev, Preprint PAR-LPTE-91-14 (1991);
E. A. Ivanov and A. A. Kapustnikov, Preprint IC-90-425 (1990);
\pl\ {\bf B267} (1991) 175.}\REF\sok{
A. Galperin   and  E. Sokatchev,  Phys.Rev.{\bf D46} 714, (1992);
	 F.   Delduc,  A. Galperin,  and   E.  Sokatchev,
	Nucl.Phys. {\bf B368} 143, (1992);
	F.   Delduc,  A. Galperin, P. Howe and   E.  Sokatchev,
	Phys. Rev. {\bf D47} (1993) 578.}\REF\berk{N. Berkovits,
Stony Brook preprint ITP-SB-92-42;
King's College preprints KCL-TH-92-6, KCL-TH-93-3;
Nucl. Phys. {\bf B379} 96, (1992);
Nucl. Phys. {\bf B358} 169, (1991);
Nucl. Phys. {\bf B350} 193, (1991).}There are also a number of
other approaches [\twistor,\sok,\berk], based on either
reformulating the superparticle in terms of twistor variables  or
using harmonic superspace (which also involves the introduction
of an infinite
number of extra variables); these will not be discussed here.

We will
consider the quantization of these models
and
   analyse   their spectra
     using  several different approaches.
 We  will consider  light-cone gauge quantization,
 a covariant  approach in which the constraints are
 imposed on the wave-functions, and a BRST-type approach. As usual,
 the first two approaches give the same results
 for the spectrum as the analysis of the BRST cohomology class
 with zero ghost number. The full BRST approach requires
 the use of the Batalin--Vilkovisky
  formalism and requires an infinite number of ghosts
 in general. However, for some of these models,
there are two types of symmetry,
 one of which acts on both coordinates and gauge fields and the other
 of which acts only on the Lagrange multiplier gauge fields,
 reflecting the fact that the constraints are not all independent.
 Presumably it is necessary to introduce ghosts corresponding to all
 of these symmetries to obtain a complete treatment.
 However,   if ghosts are only introduced
 for the first type of symmetry, only a finite
 number are needed and it is possible
 to analyse the full BRST cohomology in this reduced formalism.
 The methods of [\GH] can be used to
 argue that the zero ghost-number
 BRST cohomology class in this reduced formalism
 is exactly the same as the zero ghost-number cohomology class
 in the full formalism with an infinite number of ghosts.
Thus to find the zero ghost-number physical states, it is sufficient
(and much simpler) to use the reduced formalism.
In the cases that have been analysed in full [\mikea],
the extra cohomology classes at higher ghost number consist
of either a single zero-momentum state, or
  copies of the zero ghost-number physical states.

Before considering models with space-time supersymmetry we will
discuss the way in which particles with spin may be described in
terms of a world-line action that incorporates either local world-line
supersymmetry or a local world-line kappa symmetry.
In section 2  particle actions will be considered that give rise to
wave-functions that are in the  vector or spinor (or higher-spin)
representations of the Lorentz group.  The vector particle
is obtained by introducing an extra anticommuting coordinate that transforms
as a Lorentz vector in such a way that the
resulting theory has $N=1$ world-line supersymmetry
\REF\vecchia{L. Brink, S. Deser, B. Zumino, P. DiVecchia and P. S. Howe,
\pl\ {\bf B64} (1976) 435.}\REF\dsfdf{F.A. Berezin and M.S. Marinov,
Ann. Phys. {\bf 104} (1977) 336.}\REF\brinka{  L. Brink, P. DiVecchia
and P. S. Howe,
\np\ {\bf B118} (1977) 76.}\REF\galvaoa{C. Galvao and T. Teitelboim,
\jmp\ {\bf 21} (1980) 1863.}\REF\penati{P. S. Howe, S. Penati, M. Pernici
and P. K. Townsend,  \pl\ {\bf B215} (1988) 555.}\REF\gershuna{  V. D.
Gershun and V. I. Tkach , JETP Lett. {\bf 29}
(1979) 320.}\REF\siegelzz{W. Siegel, \ijmp\ {\bf A3} (1988)
2713.}[\vecchia-\siegelzz].
The spinor particle
 (which is a truncation of the model
 in  [\GH ])  is obtained by introducing an extra anticommuting
 coordinate that transforms as a Lorentz spinor in such a way
that the new theory has a local kappa symmetry.  Section 2 is a
warm-up for our main objective of  constructing covariant actions
for superparticles describing the particle content of  ten-dimensional
super Yang--Mills theory. We will first need to show (in sections 3
and 4) how the $D=10$ super Yang--Mills multiplet  may be described
covariantly by either a vector or spinor superfield subject to constraints.

\REF\spinsup{S. Aoyama, J. Kowalski-Glikman, J.W. van Holten and
J. Lukierski,
\pl\ {\bf B201} (1988) 487; \pl\ {\bf B216} (1989) 133;
J. Kowalski-Glikman, \pl\ {\bf B202} (1988) 343;
E. Bergshoeff and J.  Van Holten, \pl\ {\bf B226} (1989) 93;
J.  Kowalski-Glikman and J.  Lukierski,
	     Mod.Phys.Lett. {\bf A4} 2437, (1989).}In
section 5 we will obtain the superspace generalisation
of the two types of spinning particle models considered in section 2.
(Other related forms
of spinning superparticle have been discussed
in [\twistor,\spinsup].)
These models have wave-functions that are superfields in some non-trivial
representation of the Lorentz group.  In order to incorporate constraints
such as those required to define Yang--Mills superfields in section 4 it is
necessary to add certain Lagrange multiplier terms to the superparticle
actions. As will be seen in  sections 6 and 7 this can be done in a number
of ways (some of which are equivalent to models already
in the literature and some of which are new).  This enhances the local
symmetry of the action  and upon  BRST quantization  the zero ghost-number
sector of the resulting quantum superparticles is  precisely the spectrum
of super Yang--Mills theory.

\chapter{Particles,  world-line supersymmetry and kappa symmetry.}

Before discussing particles moving in superspace, it will be useful
to consider
spinning particles moving in ordinary $D$-dimensional space-time. A bosonic
particle following a world-line $x^\mu(\tau)$ in flat space can be described
by
$$ S=\int d \tau \left( p_\mu \dot x ^\mu - {1\over 2} e p^2 \right)
	   \eqn\regtowg$$
where
$p_\mu(\tau)$
is the momentum conjugate to $x^\mu(\tau)$, $\dot x= d x/d \tau$ and $e$ is a
Lagrange multiplier imposing the constraint $p^2=0$. It is invariant under
the world-line reparameterisation
invariance
$$
\delta x^\mu=k p^\mu, \qquad \delta e = \dot k, \qquad \delta p_\mu =0.
\eqn\rtytuilkly$$
Classically,
the world-line reparameterisation invariance can be fixed by choosing the
gauge $e= constant$ giving a free theory subject to the constraint $p^2=0$.
One way of quantizing this theory is to impose the standard  commutation
relations
$$
[x^\mu, p_\nu]= i \hbar { \delta ^\mu}_\nu
\eqn\ccrs$$
and
to impose the constraint on the wave-function, \ie\ to demand that the
wave-function $\Psi(x^\mu)$ satisfies the
constraint
$$p^2 \Psi=0\eqn\brconss$$
with
$p_\mu =-i\hbar \partial _\mu$. Thus the wave-function is a scalar
field   $\Psi(x^\mu)$ satisfying the Klein-Gordon equation and the bosonic
particle is seen to describe a free scalar particle.

These results can also be obtained from a BRST approach. The ghost action
for the gauge choice $e= constant$
is
$$ \inttau \hat c \dot c\eqn\erthkjghlihjl$$
and the
ghost and anti-ghost  $c(\tau),\hat c(\tau)$ satisfy the
anti-commutation
relation
$$\{ c, \hat c \} =  \hbar \eqn\yhghhk$$
The BRST charge
is
$$Q= {1 \over 2} c p^2\eqn\brstchge$$
and the
wave-function can be taken to be a function of $x^\mu$ and the ghost
$c$:
$$\Phi(x,c)= \Psi(x)+c \Psi_1 (x)\eqn\kufeht$$
The BRST constraint
$$
Q\Phi=0\eqn\brsss$$
implies that the ghost-number zero wave-function $\Psi(x) $
satisfies \brconss. Physical states are BRST cohomology classes,
corresponding to wave-functions satisfying \brsss\ with wave-functions
differing by BRST exact pieces
identified:
$$
\Phi \sim \Phi + Q \Lambda\eqn\erthfgnhklj$$
Then \brsss\ imposes $p^2 \Psi=0$ while \erthfgnhklj\ can be used to
gauge away all $\Psi_1$ except those satisfying
$\ p^2 \Psi_1=0$.
Thus the
physical states for both ghost number zero and ghost number one are
represented by wave-functions $\Psi,\Psi_1$ satisfying the Klein-Gordon
equation $p^2 \Psi=0, \ p^2 \Psi_1=0$.

Thus the first approach gives the same results as the cohomology of
the zero ghost number sector in the BRST approach,
but the latter gives a second copy of the same spectrum at
ghost number one. This is typical of all of the systems we
will encounter in this paper;
in each case, the first approach gives the same results as the BRST
analysis of zero ghost number physical states; as
the first approach
is simpler, we
  will mostly use that in the following
   models and defer a discussion of the BRST approach to the final section.

We shall be interested in modifications of the particle action that give
wave-functions transforming in some non-trivial representation of the Lorentz
group, such as the spinor or vector representation.  All of the methods we
shall discuss for giving spin to the wave-function involve adding extra
coordinates and momenta to the $(x^\mu ,p_\nu)$ phase space.

Consider first the addition to the bosonic target space of an anticommuting
vector coordinate $\lambda ^\mu$, so that the particle world-line is
$x^\mu(\tau),\lambda^\mu(\tau)$. Suppose that the gauge-fixed particle lives
in a phase-space $(x^\mu ,p_\nu, \lambda ^\mu)$ with commutation relations
given by \ccrs\ and
$$
\{ \lambda ^\mu, \lambda ^\nu \} = \hbar \eta ^{\mu \nu}
\eqn\lamccrs$$
so
that $\lambda$ is a self-conjugate variable. A maximally anticommuting set
of the grassmann variables consists
of
\foot{$[D/2]$ denotes the integer part of $D/2$.}
$[D/2]$
independent linear combinations of the $\lambda ^\mu$, which we shall denote
by $\lambda ^\alpha$, $\alpha =1,\dots, [D/2]$. The (ghost-number zero)
wave-function can be taken to be $\Phi(x^\mu, \lambda ^\alpha)$, and the
expansion
$$\eqalign{\Phi (x^\mu ,\lambda^\alpha )
       =&\psi_{0} + \psi^{\alpha}_{1}\lambda_{\alpha}
	 +\psi^{\alpha\beta}_{2}\lambda_{\alpha}\lambda_{\beta} \cr
	&+\psi^{\alpha\beta\gamma}_{3}\lambda_{\alpha}\lambda_{\beta}
			 \lambda_{\gamma} +...
	 +\psi^{ \alpha_{1}...\alpha_{[D/2]}}_{[D/2]}
			     \lambda_{\alpha_{1}...}
			     \lambda_{\alpha_{[D/2]}}
\cr}\eqn\rthjylkfgb$$
gives
a set of $2^{\lbrack D/2\rbrack}$ component fields
$\psi_{0}(x),\ \psi^{\alpha}_{1}(x),...,\psi^{\alpha_{1}...
	       \alpha_{\lbrack D/2\rbrack}}_{\lbrack D/2\rbrack}(x)$
which can be combined into an object $\psi_{A}(\tau)$,
$(A=1,...,2^{\lbrack D/2\rbrack})$  with $2^{\lbrack D/2\rbrack}$ components.
Then $p_\mu$ and the remaining $D-[D/2]$ components of $\lambda_{\mu}$
can be represented as differential operators acting on the space with
coordinates $(x^{\mu},\lambda^{\alpha})$.
As a result, $\lambda_{\mu}$ becomes a linear operator acting on $\psi_A$
which can be represented as a matrix $ \Gamma^{ \ \ B}_{\mu A}$.
Then \lamccrs\ implies that the matrices $\Gamma^{ \ \ B}_{\mu A}$ can be
normalised so as to satisfy the Clifford
algebra
$$\lbrace\Gamma_{\mu},\Gamma_{\nu}\rbrace=2\eta_{\mu\nu}\eqn\cliffo$$
so
that the $ \Gamma^{  \ \ B}_{\mu A}$ are gamma matrices for the Lorentz group
$SO(D-1,1)$. The
operators
$$L_{\mu\nu}=i(x_{\mu}p_{\nu}-x_{\nu}p_{\mu})
	     +{1\over4}\left\lbrack\lambda_{\mu},\lambda_{\nu}\right\rbrack
\eqn\lorop$$
generate
the $SO(D-1,1)$ Lorentz algebra and $\psi_A$ transforms as a spinor under
this.
Thus introducing $\lambda ^\mu$ has led to a spinor wave-function as required.

The Dirac equation for this wave-function is
$p^{\mu}\Gamma^{\ \ B}_{\mu A}\psi_{B}=0$
or
$p^{\mu}\lambda_{\mu}\Phi=0$, and this would arise if the classical particle
were subject to the constraint $p^{\mu}\lambda_{\mu}\ =0$. Squaring this
constraint gives $p^{2}=0$ and so this is needed to close the algebra of
(first-class) constraints. This motivates the classical
action
$$S=\int d\tau\left(p_{\mu}\mathop{\dot{x}}\nolimits^{\mu}
    +{i\over2}\lambda^{\mu} \dot{\lambda} ^{\mu}
    -{1\over2}ep^{2}
    +i\chi p_{\mu}\lambda^{\mu}\right)
\eqn\wlsus$$
where $e,\chi$ are lagrange multipliers imposing the
constraints
$$p^{\mu}\lambda_{\mu}\ =0,  \qquad
		   p^{2}=0.\eqn\ghdfh$$
This
is in fact the action for a
superparticle with local world-line supersymmetry and $\chi$ is the
world-line gravitino [\vecchia ,\penati].
The gauge choice $e=constant,\  \chi=0$ leads to the free theory with
commutation relations \ccrs\ and \lamccrs, subject to constraints
corresponding to \ghdfh. A BRST analysis  leads to a wave-function
$\Psi=\Phi(x^{\mu},\lambda_{\alpha}) +\gamma$ where $\gamma$ denotes
ghost-dependent terms and the BRST constraint  $Q\Psi=0$  imposes the
constraints
$$p^{\mu}\lambda_{\mu}\Phi=0,\qquad p^{2}\Phi=0\eqn\rthyklht$$
so
that the ghost-number zero physical states correspond  precisely to a
Dirac particle. For further details of this construction, see
[\vecchia ,\penati ].

An alternative way to obtain a spinor wave-function is to introduce an
extra spinor coordinate $\phi^{A  }$, together with its conjugate momentum
${\hat{\phi}} _{A}$, so that the phase space is $(x,p,\phi, \hat \phi)$.
We choose the commutation relations given by \ccrs\
and
$$\lbrace\phi^{A}, {\hat{\phi}} _{B}\rbrace
		  =\hbar \delta^{A}_{B}\eqn\teyty$$
if
$\phi, \hat \phi$ are grassmann-odd,
or
$$\lbrack\phi^{A}, {\hat{\phi}} _{B}\rbrack=i\hbar\delta ^{\ A}_{B}
\eqn\rtytr$$
if
they are commuting. Here ${\hat{\phi}} _{A}$ transforms as the conjugate
spinor representation, so ${\hat{\phi}} _{A}\phi^{A}$ is a Lorentz singlet.
Then the wave-function can be taken to
be
$$\Psi(x^{\mu},\phi^{A})=\psi_{0}(x)+\psi _{1A}(x)\phi^{A}
			 +\psi_{2AB}(x)\phi^{A}\phi^{B}+...\eqn\tetgdfg$$
and
we see that $\psi _{1A}(x)$ is a spinor wave-function. The constraint
$(N-1)\Psi=0$ where the number operator is $N=\phi^{A} {\hat{\phi}} _{A}$
implies $\Psi=\psi _{1A}(x)\phi^{A}$, so that only the spinor wave-function
survives. Further
imposing
$$p^{2}\Psi=0, \qquad
  p^{\mu}\mathop{(\Gamma}\nolimits_{\mu})_{A}^{\ B} {\hat{\phi}}_{B}\Psi=0
\eqn\yuiloru$$
implies
that $\psi_1^A$ satisfies the Dirac
equation
$$p^{\mu}\mathop{(\Gamma}\nolimits_{\mu})_{A}^{\ B}\psi _{1B}=0\eqn\tryerty$$
These
constraints   arise from the classical
action
$$S=S_1+S_2\eqn\tyugy$$
where
$$S_{1}=\int d\tau\left(p_{\mu} {\dot{x}} ^{\mu}
	 +{i}{\hat{\phi}} _{A} {\dot{\phi}} ^{A}-{1\over2}ep^{2}
	 +i\chi p_{\mu}\Gamma^{\mu}\hat{\phi}\right)
\eqn\hdfgh$$
$$S_{2}=\int d\tau\lambda(N-1)\eqn\rtygrh$$
and
$e,\chi^A,\lambda$ are lagrange multipliers imposing constraints. In
particular, $\chi^{A}$ is a gauge field for a kappa-type symmetry
$$\delta\phi=\psh \kappa , \qquad
  \delta\chi=\dot{\kappa}-i \lambda\kappa , \qquad
  \delta x^{\mu}=i\hat{\phi}\Gamma^{\mu}\kappa
\eqn\rtyeyfg$$
The
action \hdfgh\ becomes Siegel's original superparticle [\siege] after a
field redefinition if $\phi$ is anticommuting (identifying
$\phi^A \sim\theta^ A$) so we see that a spinning (Dirac) particle is
obtained by adding the lagrange multipler term \rtygrh\ to the Siegel
superparticle \hdfgh. The action \hdfgh\ has an $N=2$ twisted
space-time supersymmetry [\mikea] which is completely
broken by the term \rtygrh.

An alternative action
(generalisations of which will be relevant later) is given
by replacing \rtygrh\
with
$$S_{2}={1\over2}\inttau \lambda^{AB}\hat{\phi}_{A}\hat{\phi}_{B}
\eqn\grhull$$
leading
to the constraint $ \hat{\phi}_{A}\hat{\phi}_{B}\Psi=0$, implying
$\Psi =\psi_{0} +\psi _{1A} \phi^{A}$. This leads to a spectrum that is
reducible, described by a scalar wave-function $\psi_0$ satisfying the
Klein-Gordon equation and a spinor wave-function $\psi_{1A}$ satisfying the
Dirac equation.

A variation of this is to consider
$$S_{2}={1\over2}\inttau \lambda^{AB}d_{A}d_{B}
\eqn\grhulla$$
instead of \grhull, where
$$
d_{A}=\hat{\phi}_{A} +p_\mu (\Gamma ^ \mu)_{AB} \phi ^B
\eqn\asdtrwe$$
If $\phi$ is anticommuting then (identifying
$\phi^A \sim\theta^ A$ and making a field redefinition)
this becomes
the modified superparticle given in [\sieggya,\siegeb].
This leads to the wave-function \tetgdfg\
satisfying the constraint
$$
d_{A}d_{B} \Psi=0
\eqn\gtioeriouog$$
and this implies that $\Psi$ must be constant, representing a
zero-momentum state [\GH]. This constraint breaks the twisted
$N=2$ supersymmetry down to $N=1$.
(An alternative treatment is given in
[\MikS,\KVanPT].)

To summarise, we have seen that a spinor wave-function can be obtained
either from a spinning particle  \wlsus\ with local world-line supersymmetry,
or from a particle action \tyugy-\rtygrh\ with local kappa symmetry
\rtyeyfg. In the latter case, the extra coordinate $\phi$ can be either
anticommuting (in which case the action is closely related to Siegel's
action [\siege]) or commuting.
\REF\penn{P. S. Howe, S. Penati, M. Pernici and P. K. Townsend,
Class. Quantum Grav. {\bf 6} (1989) 1125.}We now turn to the
construction of particles with wavefunctions
that are other
non-trivial representations of the Lorentz group. One approach is to
generalise the model \tyugy-\rtygrh\ with $N=1$ local world-line
supersymmetry to one with $N$ extended local world-line supersymmetry. In
one version of this model [\siegelzz,\penn], there are $N$ anticommuting
vectors
$\lambda ^\mu_i$, ($i=1,\dots ,N$) and the corresponding wave-function is a
multi-spinor $\Psi_{A_1 A_2 \dots A_N}$ [\siegelzz,\penn].

Another approach is to modify the model \tyugy-\rtygrh\ by replacing the
spinor variables $\phi, \hat \phi$ by a coordinate $\phi_R$ in some
representation $R$ of the Lorentz group with a conjugate momentum
$\hat \phi _{\bar R}$ in a conjugate representation $\bar R$ (so that
$R \otimes \bar R$ contains a Lorentz singlet). In this way, expanding the
wave-function $\Phi(x, \phi)$ in $\phi$ gives
$\Phi = \psi _0 + \psi _1 \cdot \phi+ \dots$ where $\psi _1(x)$ is a
wave-function in the $\bar R$ representation. Imposing the constraint
$(N-1)\Phi=0$ where $N$ is the number operator $ \phi \cdot \bar
\phi$ then
gives  $\Phi =   \psi _1 \cdot \phi $ as required. One then imposes
$p^2 \Phi=0$, possibly  together with other $p$-dependent constraints,
generalising the constraint $\psh \hat \phi\Phi = 0$ of the model
\tyugy-\rtygrh. For example, consider the case in which $R$ is the vector
representation, so that the phase space has coordinates
$x^\mu, p_\mu, \phi_\mu, \hat \phi ^\mu$. We shall suppose that
$\phi, \hat \phi$ are commuting; all of the following analysis also holds if
they are anticommuting, with only some changes in some of the  signs. An
appropriate action is $S= S_1+ S_2$ with
$$S_{1}=\int d\tau\left(p_{\mu} {\dot{x}} ^{\mu}
	  + {\hat{\phi}} ^{\mu} {\dot{\phi}} _{\mu}
	  -{1\over2}ep^{2}
	  -\chi p_{\mu}  \hat{\phi}^\mu \right), \qquad
S_{2}=\int d\tau\lambda(N-1)\
\eqn\hdfghoi$$
where
$e,\chi ,\lambda$ are lagrange multipliers imposing the classical constraints
$p^2  =0$, $ p_\mu \hat \phi ^\mu =0$ and  $N=1$. Here $\chi $ is a gauge
field for a  symmetry corresponding to the kappa-type symmetry \rtyeyfg:
$$\delta\phi _\mu=p _\mu \kappa ,\qquad
  \delta\chi=\dot{\kappa} + \lambda\kappa, \qquad
  \delta x^{\mu }=\hat{\phi}^{\mu}\kappa
\eqn\rtyeyfgoi$$
Note
that this is closely related to the world-line diffeomorphism symmetry
\rtytuilkly. Quantization leads to a wave-function $\Phi (x, \phi)$ satisfying
the constraints
$$
p^2 \Phi=0 ,\quad p_\mu \hat \phi ^\mu \Phi =0, \quad (N-1) \Phi=0\eqn\tyueu$$
and
identified under the gauge transformation
$$
\Phi \sim \Phi + p_\mu \Lambda ^\mu\eqn\tyhld$$
This
leads to a vector wave-function $\psi_1^\mu$ satisfying
$$
p^2 \psi_1^\mu=0 ,\quad p_\mu \psi_1^\mu =0\eqn\tyueu$$
and
identified under the gauge transformations
$$\psi_1^\mu \sim \psi_1^\mu+ p^\mu   \lambda\eqn\tlyjuojyu$$
which clearly corresponds to a vector gauge potential in the Lorentz
gauge with
$D-2$ independent degrees of freedom. It is straightforward to
modify this to obtain a covariant vector particle in a general gauge.
The appropriate action is \tyugy\
with
$$S_{1}=\int d\tau\left(p_{\mu} {\dot{x}} ^{\mu}
	+{i\over2} {\hat{\phi}} ^{\mu} {\dot{\phi}} _{\mu}
	-\chi ^\nu (\eta_{\mu \nu} p^2-p_{\mu} p_\nu) \hat{\phi}^\mu \right)
\qquad S_{2}=\int d\tau\lambda(N-1),\eqn\hdfghoi$$
leading
to a wave-function $\psi_1^\mu$ satisfying
$$
p^\mu F_{\mu \nu}=0, \qquad
F_{\mu \nu}=p_\mu \psi_{1\nu }-p_\nu \psi_{1\mu}\eqn\tyuefgu$$
and
identified under \tlyjuojyu.

\chapter {Light-cone super Yang--Mills theory.}

In the next section we will obtain the constraints satisfied by
the superfield of ten-dimensional super Yang--Mills theory which
will establish the constraints to be satisfied by the superparticle
wavefunctions in the following sections.
Our method will make use of the light-cone gauge description of the
physical states
\REF\BrGS{ L. Brink, M. B. Green and J. H. Schwarz,
\np\ {\bf B223} (1983) 125.}[\BrGS ] which we will review in this section.

In the light-cone gauge  $SO(9,1)$
representations are decomposed into representations of a manifest $SO(8)$
little group. An $SO(9,1)$ vector $\amu$ $(\mu =0,1,\dots ,9)$ decomposes
into an $SO(8)$ vector $A_i$ $(i =1,\dots ,8)$ and two $SO(8)$ singlets
$\amin$, $\aplus$ $(\aplmi =\anut \pm \anine )$. A $16$-component Weyl spinor
of $SO(9,1)$, $\psiA$, $(A =1,\dots ,16)$ decomposes into two $8$-component
$SO(8)$ spinors $\psia$, $\psiadot$ ($a=1,\dots ,8$ and $\adot =1,\dots ,8$
label the inequivalent spinor representations of $SO(8)$). The $SO(8)$ gamma
matrices ${\gammai}_{a\bdot}$
satisfy\foot{ The
$SO(8)$ indices may be raised and lowered trivially using the metric
$\delta_{ij}$ and the charge conjugation matrices $C_{ab} =\delta_{ab}$
and $C_{\adot\bdot} =\delta_{\adot\bdot}$.}
$$\eqalign{&{\gammai}_{a\bdot}{\gammaj}_{\bdot b}
		      + {\gammaj}_{a\bdot}{\gammai}_{\bdot b}
			   =2\delta^{ij}\delta_{ab}  \cr
	   &{\gammai}_{\adot b}{\gammaj}_{b \bdot}
		      + {\gammaj}_{\adot b}{\gammai}_{b \bdot}
			   =2\delta^{ij}\delta_{\adot \bdot}
\cr}\eqn\algebra$$
and
we define
$$\gamma_{ij\dots k} = \gamma_{[i} \gamma_ j\dots \gamma_{k]}.\eqn\antisymm$$

The Yang--Mills multiplet consists of a gauge potential $\amu$ and a
Majorana-Weyl spinor $\lambda^A$ taking values in the Lie algebra of the
gauge group. In the light-cone gauge only the physical degrees of freedom
remain and these consist of the $8$ transverse degrees of freedom $A_i$ of
the Yang--Mills field and the $8$ spinor degrees of freedom $\lambda^{\adot}$
(the Dirac equation $\Dslash \lambda =0$ eliminates $8$ of the $16$
components of $\lambda^A$ in the light-cone gauge).  Expanding an
unconstrained superfield $\Phi (\xmu ,\theta_A )$ gives $2^{16}$ component
fields, which is clearly too large to describe the $8 + 8$ components we
require, so the Yang--Mills multiplet must be described by a constrained
superfield.

In the next section we shall find $SO(9,1)$-covariant constraints that lead
to the correct spectrum, but for now we shall content ourselves with the
$SO(8)$ covariance of the
light-cone gauge.  As a first step the superspace will be restricted to
$(\xmu ,\theta^{\adot})$, \ie, the $SO(8)$-covariant constraint
$\grad \Phi/{\grad\theta^a} =0$ will be imposed on any superfield, $\Phi$.
There are still $2^8$ component fields in $\Phi$ so further constraints
must be imposed.

One possible approach to imposing further constraints is to abandon
manifest $SO(8)$ invariance
\REF\brinkky{L. Brink, O. Lindgren and E. W. Nilsson,
			    \np\ {\bf B212} (1983) 401.}
\REF\brinkkya{S. Mandelstam, \np\ {\bf B213} (1983) 149.}and decompose the
coordinates into representations of a $U(4)$ subgroup [\brinkky,\brinkkya].
Thus $\theta^{\adot}$ is decomposed into
$\theta^\alpha$, ${\bar\theta}_\alpha$ which transform as a $4$ and $\bar 4$,
respectively. Then imposing the extra constraints
$\grad \Phi/{\grad\bar \theta_\alpha} =0$ gives a superfield
$\Phi (\xmu ,\theta^\alpha )$ which is unconstrained in the reduced
superspace with coordinates $(\xmu ,\theta^\alpha)$.
Following Siegel \REF\book{W. Siegel, {\it Introduction
to String Field Theory}, (World Scientific, 1988).}[\book ], we
shall refer to this
 as the {\it euphoric} formalism.
The superfield $\Phi (\xmu ,\theta^\alpha )$ has eight
$(1+6+1)$ bosonic and eight $(4+\bar 4)$ fermionic components and so gives
the correct spectrum for super Yang--Mills in terms of  $U(4)$-covariant
states.

Finding $SO(8)$-covariant constraints is more tricky but was solved in
[\BrGS] by choosing $\Phi$ to be either an $SO(8)$ vector superfield
$\Psii (\xmu ,\theta^{\adot})$, or
a spinor superfield $\Psia (\xmu ,\theta^{\adot})$. The vector superfield
is taken to satisfy the linear
constraint
$$(\gammai\gammaj -8\delta^{ij})_{\adot\bdot}D^{\bdot}\Psij =0,\eqn\linear$$
where
the $SO(8)$ supercovariant derivative
is
$$D_{\adot} ={\grad \over{\grad\theta^{\adot}}}
	      +\pplus\theta_{\adot}\;,\eqn\supergradd$$
and
$\pplus$ is the component of the $10$-momentum $\pmu =(\pplus ,\pmin ,p^i )$,
which is set to a constant in the light-cone gauge. Acting on \linear\ with
another $D^{\adot}$, gives the quadratic
constraint
$${1 \over{8\pplus}}D^{\adot}(\gammaij )_{\adot\bdot} D^{\bdot} \Psik
		=\delta^{jk}\Psii -\delta^{ik}\Psij ,\eqn\quadratic$$
which
was shown in [\BrGS ] to be fully equivalent to the linear constraint
\linear . It was also shown in [\BrGS ] that the constraint \linear\ (or
\quadratic ) leaves precisely the $8$ bosonic and $8$ fermionic degrees of
freedom of light-cone gauge super Yang--Mills theory.

The spinor superfield satisfies the linear constraint
[\BrGS ]
$$({\gammaijk})_{a\bdot} D^{\bdot}\Psi_a =0. \eqn\spinlin$$
Again,
this is equivalent to a quadratic constraint
$$
D_{[\adot} D_{\bdot ]}\Psi_c
	   =2\pplus\gamma^{cd}_{\adot\bdot}\Psi_d \;,\eqn\spinquad$$
where
$\gamma^{cd}_{\adot\bdot}
	= {1 \over 28}({\gammaij})_{\adot\bdot}({\gamma_{ij}})^{cd}$.
The spinor and
vector superfields are related by
$({\gammai})_{\adot b}\Psi_b ={1 \over 8} D_{\adot}\Psii$.

\chapter {Covariant constraints for super Yang--Mills theory}

We will now present $SO(9,1)$-covariant
superfield formulations of super Yang--Mills
which reduce to the $SO(8)$-covariant ones of the last section in the
light-cone gauge. We   use
 a superspace with coordinates $(\xmu, \theta_A)$,
\foot{Upper  and lower $SO(9,1)$ Weyl spinor indices
$A=1,2,\dots ,16$ are used to distinguish chirality, so that $\Phia$ and
$\thetaA$ ($A=1,2,\dots ,16$) have opposite chirality. We use a Majorana
representation in which the gamma matrices, $\Gamma_{\mu AB}$ and
$\Gamma_\mu ^{AB}$, are real and symmetric and satisfy
${\Gamma_\mu}_{AC}{\Gamma_\nu}^{CB}
  +{\Gamma_\nu}_{AC}{\Gamma_\mu}^{CB}= 2\delta^{\ B}_A \eta_{\mu\nu}$.}
and supercovariant derivatives
$$D^A ={\grad \over{\grad\theta_A}} + \psh^{AB}\theta_B .\eqn\supergrad$$

We will now show that the  $SO(8)$ vector superfield
$\Psi_i (\xmu ,\theta^{\adot})$ satisfying \linear\ may be obtained from an
$SO(9,1)$ vector superfield $\Psi_\nu (\xmu ,\thetaA )$ together with the
linear constraints
$$ p^2 \Psimu =0, \qquad ({\psh}_{AB} D^B )\Psimu =0,\qquad
				\pmu\Psimu =0,\eqn\onelinear$$
and
$$ D^A \Psi^\mu
	 =\seventh (\Gamma^{\mu\nu}{)^A}_B D^B\Psi_\nu\;,\eqn\twolinear$$
where
$\Gamma_{\mu\nu\dots\rho} =\Gamma_{[\mu}\Gamma_\nu \dots\Gamma_{\rho]}$.
The first constraint in \onelinear\ implies the momentum is null and the
remaining
constraints will be analysed in a Lorentz frame in which the
momentum is $\pmu =(\pplus ,0,\vec 0 \;)$, \ie, $\pmin =0$, $p^i =0$.

It will be convenient to define $m^\mu =\pmu /\pplus$ and introduce a null
vector $n^\mu$ such that $m^2 =0$, $n^2 =0$ and $m\cdot n =1$,
and then choose a representation of the gamma matrices such that the
projectors ${\msh}{\nsh}$ and ${\nsh}{\msh}$ are diagonal. The Lorentz
transformations preserving the null vectors $m^\mu, n^\mu$ form an $SO(8)$
transverse group and any chiral $SO(9,1)$ spinor, $\chi^A$ or $\psi_A$,
may be decomposed into two inequivalent $SO(8)$ spinors
(distinguished by dotted and undotted indices, $\adot$ and $a$)
as
$$\psi_{A}=\left(\matrix {\psi_{a} \cr\noalign{\smallskip}
			  \psi_{\adot} \cr}\right) \; \qquad
  \chi^{A}=\left(\matrix {\chi_{a} \cr\noalign{\smallskip}
			  \chi_{\adot} \cr}\right)
\;.\eqn\ryhgk$$
In this basis,
the matrices $m \cdot  \Gamma $ and $n \cdot  \Gamma $
take the form
$$\eqalign{
({\msh} )_{AB}&=\left(
		\matrix{    0              &      0  \cr \noalign{\smallskip}
			-\delta_{\adot b}  &      0  \cr}\right)
\;, \qquad
({\nsh} )^{AB}= \left(
		\matrix{    0  &-\delta_{a\dot{b}}   \cr \noalign{\smallskip}
			    0              &      0  \cr}\right)
\;,\cr
({\msh} )^{AB}&= \left(
		\matrix{    0              &      0  \cr \noalign{\smallskip}
			 \delta_{\dot{a}b} &      0  \cr}\right)
\;,\qquad
({\nsh} )_{AB}= \left(
		\matrix{    0   &  \delta_{a\bdot}   \cr \noalign{\smallskip}
			    0   &      0             \cr}\right)
\;.\cr}\eqn\tryee$$
Spinors $\chi^A$, $\psi_A$ can be decomposed as follows
$$\eqalign{&
	     ({\msh} )_{AB} \chi^B \rightarrow - \chi_a    \;,\quad
	     ({\nsh} )_{AB} \chi^B \rightarrow \chi_{\adot} \;,\cr
	   &
	     ({\msh} )^{AB}\psi_B \rightarrow \psi_{\adot} \;,\quad
	     ({\nsh} )^{AB}\psi_B \rightarrow - \psi_a    \;.
\cr}\eqn\decompose$$

It follows that
${\psh}_{AB} D^B \rightarrow -\pplus\grad /{\grad\theta^a}$
and
the second constraint in \onelinear\
becomes
$${\grad \over{\grad\theta^a}}\Psimu =0,\eqn\newonelin$$
while
$\pmu\Psimu =p^+ \Psiplus=0$ implies that $\Psiplus =0$ if $p^+ \ne0$.
We are then left with superfields $\Psimin (\xmu ,\theta^{\adot})$
and $\Psi_i (\xmu ,\theta^{\adot})$ satisfying \twolinear , which
implies
$$D^{\adot}\Psimin =0.\eqn\leftlinear$$
Using
$D_{\adot} ={\grad /\grad\theta^{\adot}} +\pplus\theta_{\adot}$,
we find that \leftlinear\ implies
$D^{\adot} D^{\bdot} \Psimin =2\pplus  \delta ^{\adot \bdot}\Psimin=0$
  and hence
$\Psimin =0$, if $\pplus \not=0$. This leaves a superfield
$\Psi_i (\xmu ,\theta^{\adot})$
satisfying
$$ D^{\adot} \Psii =\seventh (\gammaij )^{\adot\bdot} D_{\bdot}\Psi_j
							\;,\eqn\superleft$$
which
is precisely the super Yang--Mills constraint \linear . Thus the covariant
constraints \onelinear\ and \twolinear\ give a covariant superfield
formulation of super Yang--Mills.

Similarly, the quadratic constraint \quadratic\ is recovered from the
following covariant constraints
$$p^2 \Psimu =0, \qquad ({\psh}_{AB} D^B ) \Psimu =0, \qquad
				      \pmu \Psimu =0, \eqn\onequadr$$
and
$$D^A D^B \Psimu
    + 8 (\Gamma_\mu )^{C[A} (\Gamma^\nu\psh {)_C}^{B]} \Psi_\nu =0.
							   \eqn\twoquadr$$
To see
that this is the correct quadratic covariant formulation of super Yang--Mills,
we follow a null frame analysis similar to the above and set
$\pmu =(\pplus ,0,\vec 0 \;)$. The last equation in \onequadr\ implies that
$\Psiplus =0$ while  ${\psh}_{AB} D^B \Psimu =0$ implies that
$\Psimu =\Psimu (x ,\theta^{\adot} )$.
We are then left with superfields $\Psimin (\xmu ,\theta^{\adot})$
and $\Psi_i =\Psi_i (\xmu ,\theta^{\adot})$ satisfying \twoquadr .
This constraint implies
$\Psimin =0$ (provided that $\pplus \not= 0$) leaving a superfield $\Psi_i$
satisfying
$$D^{\adot} D^{\bdot} \Psii =
	 8\pplus (\gammai )^{c[\adot} (\gamma^j {)_c}^{\bdot ]}
			     \Psi_j \; \eqn\conequadr$$
which is
equivalent to \quadratic .

The spinor constraints can   be obtained similarly.
 The spinor superfield
$\Psi_a (\xmu ,\theta^{\adot})$ must come from an $SO (9,1)$ spinor
superfield $\PsiA (\xmu ,\thetaA )$. We impose the linear
constraints
$$p^2 \PsiA =0, \qquad ({\psh}_{AC} D^C ) \PsiB =0, \qquad
			      {\psh}^{AB} \PsiB =0, \eqn\onenice$$
and
$$(\Gamma^{\mu\nu\rho\sigma}{)_A}^B D^A \PsiB =0,\eqn\twonice$$
where
$\Gamma^{\mu\nu\rho\sigma}
     = \Gamma^{[\mu}\Gamma^\nu \Gamma^\rho \Gamma^{\sigma ]}$.
The first
constraint in \onenice\ implies the momentum is null so that the remaining
ones can be solved in the Lorentz frame in which $\pmu =(\pplus ,0,\vec 0 )$.
The second constraint in \onenice\ again implies that
$\PsiB =\PsiB (\xmu ,\theta^{\adot})$. The last constraint
in \onenice\ becomes $\pplus\Psi_{\adot}=0$ and hence $\Psi_{\adot}=0$
(provided that $\pplus \not= 0$), leaving a superfield $\Psi_a$
satisfying
$$(\gamma^{ijk})^{a \bdot} D_{\bdot}\Psi_a =0, \eqn\conspinquad$$
which
is precisely the light-cone Yang--Mills constraint \spinlin .

Finally, the quadratic spinor constraint \spinquad\ can be obtained
from
$$p^2 \PsiA = 0, \qquad ({\psh}_{AC} D^C ) \PsiB =0, \qquad
			       {\psh}^{AB} \PsiB =0, \eqn\onenices$$
and
$$D^A D^B \PsiC
     + 8 (\Gamma^{\mu})^{D[A}(\Gamma_{\mu}\psh {)^{B]}}_C \Psi_D =0.
						     \eqn\twonices$$
The second constraint in \onenices\ implies that
$\PsiB =\PsiB (\xmu ,\theta^{\adot})$, if $\pplus \not= 0$. The last
constraint in \onenices\ becomes $\pplus\Psi_{\adot} =0$ and hence
$\Psi_{\adot} =0$ while the former expresses that the momentum is null.
We are then left with a superfield $\Psi_a$
satisfying
$$D^{\adot} D^{\bdot} \Psi_c
	   -8\pplus (\gammaij )^{\adot\bdot}
		    (\gamma_{ij} )_{cd} \Psi^d =0.\eqn\qqconespin$$
which is \spinquad .

\chapter {Covariant superparticle actions.}

The remaining sections of this paper are aimed at generalizing the
particle actions of section 2 to incorporate space-time supersymmetry
in such a manner as to reproduce covariant  super Yang--Mills
wave--functions.  This section will be concerned with the superspace
generalization of section 2 ignoring the issue of the Yang--Mills
constraints, which will be dealt with in the next two sections.

  \section {The original  superparticle.}

To begin with it will be useful to review the superparticle of
[\casal,\BrSch ] even though there appears to be no way of quantizing
it in a manifestly covariant manner.  It is defined by the
action
$$S_{0} =\inttau \Bigl[ p_\mu ({\dot x}^\mu -i\bar\theta\Gammu\dot\theta )
	    - \half e p^2 \Bigr],\eqn\BSCaction$$
where
$\dot\theta =d\theta /d\tau$.
This describes a particle with world-line parametrized by $\tau$
moving through a $10$-dimensional $N=1$ superspace with coordinates
$(\xmu ,\theta_A )$. The momentum of the particle is $\pmu$ and
$e$ is a world-line einbein.
The action $S_{BSC}$ is invariant under a $10$-dimensional
super-Poincar\'e symmetry
$$\delta \theta =\epsilon \;,\qquad
  \delta \xmu =i\epsilon \Gammu \theta \;,\eqn\poincare$$
(where
$\epsilon_A$ is a constant Grassmann parameter)
and it is invariant also under world line-reparametrizations together
with the local kappa
symmetry
$$\eqalign{& \delta \theta =\psh \kappa\;, \qquad
	     \delta e =4i\kappa\dot\theta +\dot\xi\;,       \cr
	   & \delta \xmu =i\theta\Gammu\psh\kappa + \xi\pmu \;,
	     \qquad \delta \pmu =0.
	   \cr}\eqn\localsym$$
The Grassmann spinor
$\kappa_A$ parametrizes the local fermionic symmetry, while $\xi$
parametrizes a linear combination of world-line diffeomorphisms and a
\lq trivial' local symmetry.
\foot{A \lq trivial' symmetry is one under
which all fields transform into equations of motion,
 so that the symmetry
       does not eliminate on-shell
      degrees of freedom  Any action $S(\phi ^i)$
dependent on fields $ \phi ^i$ will automatically be invariant under
local transformations of the
form $\delta \phi ^i=\lambda J^{ij}(\phi)\delta S
/\delta \phi ^j$ (with local parameter $\lambda$) provided $J^{ij}$ is
(graded) anti-symmetric. The
corresponding Noether current vanishes on-shell
\REF\romans{L. J. Romans, \np\ {\bf B281} (1987) 639.}
\REF\Auria{P. K. Townsend, in {\it Superunification and Extra Dimensions},
	   eds. R. D'Auria and P. Fr\'e,
(World Scientific, 1986).}[\romans,\Auria ].}

These symmetries can be fixed by going to the light-cone gauge,
in which the reparametrization invariance is used to set $e$ to be
a constant and the fermionic symmetry is used to impose the condition
$\gamma^{+}\theta = 0$, eliminating half of the components of $\thetaA$.
Finally, the condition $x^+=p^+ \tau +x_0^+$
is imposed for some constants
$p^+ ,x_0^+$, and the constraint $p^2=0$ is solved to give $p^-=p^ip^i/p^+$.

The light-cone action is
$$S_{lc} =\inttau \Bigl(p^i{\dot x}^i -\half {p^i}{p^i}
			+\half i \theta^{\adot}{\dot\theta}^{\adot}
			\Big), \eqn\Slight$$
(where factors of $\pplus$
have been absorbed into redefinitions of the fields) which is a free action.
In an operator approach it is quantized by imposing the  (anti-) commutation
relations
$$[x^i ,p^j ]=-i\delta^{ij}\;, \qquad
  \{ {\theta}^{\adot},{\theta}^{\bdot}\} =2\delta^{\adot\bdot}.
\eqn\commrel$$

It is convenient to use a  euphoric ($U(4)$) basis, in which $\theta^{\adot}$
is written in terms of $\theta^\alpha$ ($\alpha=1,\dots, 4$)
and its complex conjugate  ${\bar\theta}_{\alpha}$, so that the kinetic term
$\half i\theta^{\adot}{\dot\theta}^{\adot}$ becomes
$i{\bar\theta}_{\alpha}{\dot\theta}^\alpha$ and the anti-commutation
relations become
$$\{ {\bar\theta}_\alpha ,\theta^\beta \}
	     = {\delta_\alpha}^\beta .\eqn\anticomm$$

The wave-function is a function of a maximal commuting
subset of the phase space variables, and these can be taken to be
$(p^i ,\pplus ,\theta^\alpha )$.
\foot{Recall that in the light-cone gauge, $\pmin =p^i p^i /\pplus$.}
It is then an unconstrained function $\Psi (p^\mu ,\theta^\alpha )$ satisfying
$p^2 \Psi =0$. Expanding $\Psi$ in $\theta^\alpha$ gives the eight bosonic
and eight fermionic component fields of super Yang--Mills.
Since this theory cannot be covariantly quantized we now turn to consider
other superparticle actions
which lead to the covariant constraints described in the last section.

 \section{Twisted superparticles}

In [\siege,\siegea],
Siegel proposed the following modified superparticle
action:
$$S_{t} =\inttau \Bigl[ p_\mu ({\dot x}^\mu
		-i\bar\theta\Gammu\dot\theta ) + id \dot\theta
		- \half e p^2 + i \psi \psh d \Bigr],\eqn\twist$$
where
$d$ is introduced so that $\theta$ has a  conjugate momentum
${\hat\theta}^A =d^A -{\psh}^{AB}\thetaB$, while $\psi$ is a gauge field
for the kappa symmetry (which we shall sometimes refer to as
the {\cal B} symmetry)
$$\eqalign{&
	\delta \psi = \dot \kappa, \qquad \delta \theta = \psh \kappa,
\qquad
	\delta d= 2 p^2 \kappa,\cr
      & \delta x^\mu = -i \kappa \psh \Gamma^\mu \theta  ,\qquad
	\delta e = 4i \dot \theta \kappa
.\cr}\eqn\twikap$$
The action is also invariant under world-line reparameterisations
(the {\cal A} symmetry)
and the  {\cal E} and {\cal F} symmetries given by
$$ \delta\psi =  \pslash
\eta , \quad \delta e = - 2 i\bar d \eta .\eqn\syme$$
and
$$
\delta \psi^A = \omega d^A
\eqn\sillysym$$

The
action \twist\ was shown in [\mikea ] to be invariant under an $N=2$ twisted
space-time super-Poincar\' e symmetry, which includes two supercharges
$Q_1,Q_2$ which are both Majorana-Weyl spinors in $D=10$ satisfying
$$
\{ Q_1, Q_1 \} = \psh , \qquad \{ Q_2, Q_2 \} =  -\psh , \qquad
\{ Q_1, Q_2 \} = 0 .\eqn\twistalg$$
As one might suspect from the opposite signs of the anticommutators
$\{ Q_1, Q_1 \}$ and $ \{ Q_2, Q_2 \}$  the spectrum of
the theory has negative norm states. The physical states of the
theory were shown in [\mikea ] to be represented by a scalar superfield
wave-function $\Phi(x^\mu, \theta _A) $ satisfying the
constraints
$$
p^2 \Phi=0, \qquad \psh d \Phi=0\eqn\turt$$
where
$\hat \theta $ is represented by
$\hat\theta ^A=i\hbar\partial /\partial\theta _A$ so that
$d^A= i \hbar \partial /\partial \theta _A -{\psh}^{AB}\thetaB$.

The light-cone gauge action is [\mikea]
$$S_{lc} = \int d \tau
\left ( p_i \dot x^i - {1 \over 2} p^i p^i+
i \hat \theta^a\dot \theta ^a \right),\eqn\lcact$$
where factors of $p^+$ have been absorbed into field redefinitions and
$$\pi^a = d^a - p^+\theta^a.\eqn\pidef$$
 The corresponding spectrum  contains  $2^8$ states
and is {\it not} equivalent to the $N=1$ superparticle with action \BSCaction,
which has a spectrum of $2^4$ states
\REF\allena{T. J. Allen, Mod. Phys. Lett. {\bf A2} (1987) 209.}[\allena].
Defining the $SO(8)$ spinors
$\theta ^1 = \half (\theta  + \hat \theta)$,
$\theta ^ 2=\half (\theta  - \hat \theta)$,
the light-cone action becomes
$$S_{lc} = \int d \tau \left ( p_i
\dot x^i - {1 \over 2} p^i p^i+ i\theta ^1
_a \dot \theta ^1_a -  i \theta ^ 2_a \dot \theta ^ 2_a\right).
\eqn\lcacta$$
and the relative minus sign between the two fermion
kinetic terms implies that half of the physical states
must have
negative norm.

A further modification of this theory, proposed in [\siegeb ,\sieggya ],
is to add the term
$$
 S'={1 \over 2} \inttau \chi _{AB}d^Ad^B\eqn\sspdos$$
involving
a Lagrange multiplier $\chi _{AB}$ to $S_t$ to give $S'_t = S_t + S'$.
This leads to a scalar wave-function
satisfying \turt\ together
with
$$d^A d^B \Phi=0\eqn\ieyj$$
This
constraint implies that either the wave-function vanishes, or that the
momentum $p$ is zero [\mikea ], so that the only physical states
are zero-momentum ground states.
\foot{Note that although these
      constraints would be expected to emerge from the BRST constraints for
      the ghost-number zero physical states  in any covariant BRST analysis,
      there has not yet been a complete covariant BRST analysis of this model
      and it was argued in [\mikky ] that such an analysis may not be
      possible, in the sense that there may not be any solution of the
      Batalin-Vilkovisky master equation for this system.}

\section{Spinning twisted superparticles}

In section 2 it was seen that
the usual bosonic particle had a scalar wave-function,
but by adding appropriate world-line degrees of freedom it became possible
to obtain theories with wave-functions describing  spin. We now wish to
generalise
this by adding extra degrees of freedom to the twisted superparticle
action \twist\ so as to obtain wave-functions which are superfields with
spin. As in section 2 there will be two ways of doing this, one of which
leads to an extra world-line supersymmetry and one of which leads to an
extra kappa-type symmetry.

We first generalise the  world-line supersymmetric action \wlsus\ and
consider the action
$$ S_{st}^{(1)}=
   S_{t }  -i\inttau \left( {1\over 2}\lambda ^\mu \dot \lambda ^\mu
 + \chi p_\mu \lambda ^\mu \right)\eqn\twisusy$$
with
an extra anticommuting vector coordinate $\lambda ^\mu (\tau)$, where
$S_t$ is given by \twist.
The action \twisusy\ is invariant under the local kappa
symmetry
$$\eqalign{&\delta \theta = \psh \kappa, \qquad
	    \delta \psi = \dot \kappa, \qquad
	    \delta e =4i\dot\theta\kappa \cr
	   &
	    \delta \xmu = i\theta\gammu\delta\theta + id\gammu\kappa ,
	    \qquad \delta d = 2 p^2 \kappa
,\cr}\eqn\kakakak$$
with
spinor parameter $\kappa ^A(\tau)$ and the local world-line supersymmetry
$$
  \delta x^\mu =i \epsilon \lambda ^\mu , \qquad
  \delta \chi=\dot \epsilon ,\qquad
  \delta \lambda^\mu =\epsilon \pmu\eqn\teyty$$
with
parameter $\epsilon(\tau)$.

On quantization, $\lambda^\mu $ satisfies the Clifford algebra
\lamccrs\ and the wave-function can be taken to be a scalar function
$\Phi (x^\mu ,\theta_A ,\lambda^\alpha )$ depending on $[D/2]$ of the
$\lambda$'s, $\lambda^\alpha$, $\alpha =1,\dots, [D/2]$.
The wave-function has the expansion
$$
\Phi (x^\mu ,\theta_A ,\lambda^\alpha )
	  =\psi_{0} + \psi^{\alpha}_{1} \lambda_{\alpha}
	   +\psi^{\alpha\beta}_{2} \lambda_{\alpha} \lambda_{\beta}
	   +\psi^{\alpha\beta\gamma}_{3} \lambda_{\alpha} \lambda_{\beta}
					\lambda_{\gamma} +...
\eqn\rtlkfgb$$
giving
a set of $2^{[D/2]}$ component fields
$\psi_{0}(x,\theta ),\ \psi^{\alpha}_{1}(x,\theta),...,
 \psi^{\alpha_{1}...\alpha_{[D/2]}}_{[D/2]} (x,\theta)$
which
can be combined into an object $\psi_{A}(x,\theta)$,
$ (A=1,...,2^{[D/2]})$ with $2^{[D/2]}$ components.
Then $p_\mu,\hat \theta, d$ and the remaining $D-[D/2]$ components of
$\lambda_{\mu}$ can be represented as differential operators acting on the
space with coordinates $(x^{\mu},\lambda^{\alpha})$.   In particular,
$\lambda_{\mu}$ becomes a linear operator acting on $\psi_A$ which can be
represented as a matrix $ \hat \Gamma^{\ \ \ \ B}_{\mu A}$.
Equation  \lamccrs\ implies that these matrices can be normalised so as to
satisfy the Clifford algebra \cliffo\ and the
$\hat  \Gamma^{\ \ \ \ B}_{\mu A}$ are related to the usual $SO(D-1,1)$ gamma
matrices,
 $\Gamma^{\ \ \ \ B}_{\mu A}$, by a similarity transformation.
A  basis can then be chosen in which they are identified so that
$\hat \Gamma =\Gamma$  in which case $\psi_{A}(x,\theta)$ transforms as a
spinor under the action of the modified Lorentz generators
$L'_{\mu\nu}=L_{\mu\nu} +{1\over4}\left\lbrack
		  \lambda_{\mu},\lambda_{\nu}\right\rbrack$,
where $L_{\mu\nu}$ are the standard Lorentz generators. The wave-function
$\psi$ satisfies the
constraints
$$\psh \psi =0, \qquad
   p^2 \psi =0, \qquad
   (\psh d) \psi=0\eqn\coco$$
The
spectrum is then represented by a constrained spinor superfield;
however, this corresponds to a reducible representation of supersymmetry.
Multi-spinor wave-functions $\psi_{A_1A_2 \dots A_N}$ can be obtained by
adding $N$ variables $\lambda ^\mu _i$, $i=1,\dots N$, so as to obtain
$N$ extended local world-line supersymmetry.

Next we consider the generalisation of the particle action with local kappa
symmetry defined by \tyugy,\hdfgh\ and \rtygrh\ to the superparticle action
$$ S_{st}^{(2)}=S_{t }+ S_1+S_2\eqn\pq$$
where
$$S_{1}=\int d\tau\left( {i\over2}\hat{\phi}_{A}\dot{\phi}^{A}
	 -i\chi p_{\mu}\Gamma^{\mu}\hat{\phi}\right) ,\eqn\rs$$
$$  S_{2}=\int d\tau\lambda(N-1),\eqn\tu$$
with extra spinor cordinates $\phi^A, \hat \phi _A$. The action is invariant
under the usual kappa symmetry given by \twikap\ with all other fields
invariant, and a new kappa-type symmetry given by \rtyeyfg\ with all other
fields invariant.
Instead of  \tu\ we could have equally well used \grhull\  which would
impose the constraint $\hat \phi_A \hat \phi_B =0$, as this gives rise
to the same spectrum.

The analysis of the physical states is similar to that of section 2.
The wave-function can be taken to be a function
$\Psi(x^\mu, \theta_A, \phi^A)$ and the constraint $(N-1)\Psi=0$
implies that $\Psi=\psi _A \phi^A$ for some spinor superfield
$\psi_A(x,\theta)$. The remaining constraints
are
$$ p^2\psi=0, \qquad
   (\psh d )\psi=0 ,\qquad
   \psh \psi =0. \eqn\wlsuscon$$
Again
this gives a reducible representation of supersymmetry.

Also as before, this can be generalised to obtain a wave-function in any
representation $R$ of the Lorentz group by adding a momentum
$\hat \phi$ that transforms according to the $R$ representation,
together with a conjugate coordinate $\phi$. In particular, for the vector
representation, we consider the action \pq\
with
$$S_{1}=\inttau\left(\hat{\phi}_{\mu}\dot{\phi}^{\mu}
		      -\chi p_{\mu} \hat{\phi}^\mu\right),\eqn\vw$$
and
$S_{2}$ is given by \tu\ with $N={\phi}^{\mu}\hat{\phi}_{\mu}$.
The extra vector coordinates ${\phi}^{\mu},\hat{\phi}_{\mu}$ are taken to
be commuting. The action is invariant under the standard kappa symmetry
\twikap\ together with the \lq kappa/diffeomorphism' symmetry \rtyeyfgoi.
This leads to physical states described by a vector superfield
$\psi^\mu(x,\theta)$ satisfying the constraints
$$
p^2 \psi^\mu =0, \qquad
p_\mu \psi^\mu =0, \qquad
\psh d \psi^\mu =0
\eqn\veccon$$
and
identified modulo the gauge transformations
$$
  \psi ^\mu \sim \psi ^\mu + p^\mu \lambda
  \eqn\modgag$$
for
arbitrary superfields $\lambda(x, \theta)$.
This corresponds to a super-gauge connection in Lorentz gauge and
corresponds to a reducible multiplet in general.

To summarise, there are a number of ways of constructing
superparticle actions that give rise to wave-functions that are spinor or
vector superfields
satisfying kinematic constraints, and corresponding to reducible
multiplets. It was seen in section 4 that super Yang--Mills
theory in 10 dimensions is described by precisely such wave-functions
subject to certain extra super-covariant constraints.  Superparticle
theories with spectra coinciding with  that of super Yang--Mills can
therefore be constructed by adding Lagrange
multiplier terms to these superparticle actions, leading  to
these extra constraints. The remainder of this paper is devoted to a
detailed analysis of such models.

 \chapter {Covariant superparticles with spinor super wave-functions.}

In this section we shall
obtain actions that describe a  superparticle with a spinor
super wave-function satisfying either the quadratic constraint \spinquad\ or
the
linear one \spinlin. We shall start with models with extra spinor variables
which have kappa symmetries (models of the first and second type)
and then consider models with extra vector coordinates (models of the
first ilk) which have world-line supersymmetry. The models with the
quadratic constraint of the first type and the first ilk were originally
presented in [\GH ] and [\rocek ], respectively.

 \section {First type (quadratic constraint).}

It was shown in the last section that the model  \pq-\tu\ in which an extra
spinor coordinate $\phiA$ was introduced, together with its conjugate
momentum $\hatphiA$ and a momentum ${\hat\theta}^A$ conjugate to $\thetaA$,
gives a spinor wave-function superfield subject to the constraints \wlsuscon\
(recall that ${\hat\theta}^A =d^A -{\psh}^{AB}\thetaB$).  To
describe super Yang--Mills we wish to impose the extra constraint
\spinquad\ which  can be done by adding to the action \pq-\tu\
the  term
$$S_3 =\inttau \Bigl[ \half d\chi d
		     +2 \hat\phi\Gammu\chi\Gamma_\mu \psh\phi
					  \Bigr] \eqn\zeroaction$$
involving a Lagrange multiplier $\chi_{AB}=-\chi_{BA}$.
The alternative action in which the
constraint $N=1$ imposed by \rtygrh\  is replaced by the constraint
$\hat\phi\hat\phi=0$ that comes from the term \grhull\ (as in  [\GH ])
gives a very similar model so here we shall simply review  the model of [\GH].
The total action is the sum of a free action
\foot{Spinor indices are supressed and we use a matrix notation, so that
	    $d\dot\theta =d^A {\dot\theta}_A$ ,
	   $\theta\Gammu{\dot\theta} =\thetaA{(\Gammu )}^{AB}{\dot\theta}_B$,
	   $d\chi d =d^A {\chi_{AB}} d^B$,
	   $\hat\phi\Gammu\chi\Gamma_\mu\psh\phi =
		 {\hat\phi}_A{(\Gammu )}^{AB}\chi_{BC}
		       {(\Gamma_\mu )}^{CD}{\psh}_{DE}\phi^E$, etc.}
$$
  S_0 =\inttau \Bigl[p_\mu {\dot x}^\mu
	 + i\hat\theta\dot\theta
	 +i\hat\phi\dot\phi \Bigr],\eqn\zeroaction$$
plus the
term
$$S' =\inttau \Bigl[-\half e p^2 + i\psi\psh d
		     + i\Lambda\psh\hatphi
		     + \half\hat\phi\Upsilon\hat\phi +\half d\chi d
		     +2 \hat\phi\Gammu\chi\Gamma_\mu \psh\phi
					  \Bigr],\eqn\oneaction$$
where
$e$, $\psi^A$, $\chi_{AB}=-\chi_{BA}$, $\Lambda_A$ and
$\Upsilon^{AB} =-\Upsilon^{BA}$ are Lagrange multipliers imposing
the first class
constraints
$$\eqalign{& p^2 =0 \;, \qquad \psh d =0 \;, \qquad \psh\hat\phi =0 \;,\cr
	     {\hat\phi}_A{\hat\phi}_B &=0 \;, \qquad
	     d^A d^B -8(\hat\phi \Gammu )^{[A} (\Gamma_\mu \psh\phi )^{B]} =0
,\cr}\eqn\conditions$$
and
are also gauge fields for corresponding local symmetries.

The action  is invariant under  the global space-time supersymmetry
transformations,
$$
  \delta \theta= \epsilon,\qquad
  \delta x^\mu=i\epsilon\Gamma^\mu\theta ,\eqn\spacetime$$
(where
$\epsilon_A$ is a constant Grassmann parameter) and  a number of
local symmetries, which generalize ones found for the earlier superparticle
actions.
The symmetries divide into two kinds [\KVanPT].
Symmetries of the `first kind' are those under which a gauge field transforms
into the derivative of a gauge parameter.
These include world-line reparameterizations which, when
 combined with a   trivial  symmetry,
gives the  $\cal A$ transformations
$$\delta x^\mu = \xi p^\mu,  \qquad \delta e = \dot \xi,\eqn\asymm$$
the other fields being inert.
There are also two fermionic symmetries of the first kind, $\cal B$ and
$\cal B'$, with
fermionic spinor parameters $\kappa^A (\tau)$ and $\zeta_A (\tau)$,
$$
  \delta \theta = \psh \kappa ,\quad
  \delta\psi = \dot \kappa ,\quad
  \delta x^\mu = i\theta\Gamma^\mu\psh\kappa + id\Gamma^\mu\kappa ,\quad
  \delta e = 4i\dot \theta\kappa
,\eqn\btrans$$
$$
  \delta\phi =\zeta ,\qquad
  \delta \Lambda = \dot\zeta ,\qquad
  \delta x^\mu = i\hat\phi\Gamma^\mu\zeta ,\qquad
  \delta e =4\hat\phi \Gamma^\mu\chi\Gamma_\mu\zeta
.\eqn\bprime$$
Finally, there are further bosonic symmetries of the first kind associated
with the  gauge fields $\chi,\Upsilon$
(the $\cal C$ and ${\cal C}'$ symmetries)   defined by
$$\eqalign{&
     \delta\theta = -i\rho d, \qquad
     \delta d = -2i  \rho d, \qquad
     \delta\chi = \dot \rho + 2i (\chi\psh\rho-\rho\psh\chi),\cr
	   &
     \delta x^\mu =i\theta\Gamma^\mu\delta \theta
		   -2i \hat\phi\Gamma^\nu\rho\Gamma_\nu \Gamma^\mu\phi ,\qquad
     \delta e = 4\psi \rho d
	       + 4 \Lambda\Gamma_\mu \rho \Gamma_\mu \hat \phi ,\cr
	   &
     \delta \phi = 2i \Gamma_\mu\rho\Gamma^\mu\psh \phi ,\qquad
     \delta \hat\phi = - 2i \psh\Gamma_\mu\rho\Gamma_\mu \hat\phi ,\cr
	   &
     \delta \Upsilon = 2i (\Upsilon \psh \Gamma_\mu\rho\Gamma^\mu
		 + \Gamma_\mu\rho\Gamma^\mu \psh \Upsilon ), \cr}
\eqn\ctrans$$
$$
  \delta \Upsilon =\dot v -2i \Gamma^\mu\chi\Gamma_\mu \psh v
	      +2i v \psh \Gamma_\mu \chi \Gamma^\mu ,\qquad
  \delta\phi = iv \hat \phi
,\eqn\cprime$$
where
$\rho_{AB}=-\rho_{BA}$ and $v^{AB}=-v^{BA}$ are bosonic bispinor
parameters.

There are also local
symmetries of the `second kind' that act only on the gauge fields.
These are the $\cal E$ and $\cal E'$ symmetries with fermionic parameters
$\eta_A$ and $\omega^A$
$$\delta\psi =\psh \eta, \qquad
  \delta \Lambda = \psh \omega , \qquad
  \delta e = -2i d\eta -2i \hat \phi \omega,
    \eqn\etrans$$
and
the $\cal F$,  $\cal F'$ and $\cal G'$ symmetries with  parameters
$\sigma^A_{\ B}$, $u_A^{\ B}$ and $\Sigma^{ABC}$, respectively, which are
given by
$$\eqalign{&\delta\psi = \sigma d, \qquad
	    \delta\chi = i(\psh \sigma -\sigma ^t \psh) ,\qquad
	    \delta \Lambda = u\hat \phi - 2 (\Gamma_\mu \sigma \Gamma^\mu
					+ 2 \sigma^t)\psh \phi ,\cr &
	    \delta \Upsilon ^{AB} = 2i\psh^{C[A}{ u_C} ^{B]}
			       + \Sigma^{ABC}\hat\phi_C ,\qquad
	    \delta e = 4i\hat \phi (2\sigma
		       + \Gamma_\mu \sigma^t \Gamma^\mu )\phi ,\cr}
\eqn\ftrans$$
where $\Sigma^{ABC}=-\Sigma^{BAC}$,  $\Sigma^{[ABC]}=0$ and
the transpose of  $\sigma^A_{\ B}$ is $(\sigma^t)_B^{\ A}$.
The presence of the symmetries of the second kind reflects ambiguities
in the definition of the $\cal A$, $\cal B$, $\cal B'$, $\cal C$ and
${\cal C}'$ symmetries and relations between the corresponding constraints.

The covariant quantization of this superparticle was  discussed
in [\GH ] in the gauge $e=1$ with the other gauge fields set
to zero.
Covariant quantization requires  the methods of Batalin and Vilkovisky  [\BV]
since the gauge algebra only closes on shell, and requires an infinite
number of ghost fields since the symmetries are infinitely reducible.
The classical gauge-fixed action takes the
form
$$S_{fixed} =\inttau \Bigl[ p_\mu {\dot x}^\mu -\half p^2
		  +i\hat\theta\dot\theta
			   +i\hat\phi\dot\phi  \Bigr]\eqn\fixedact$$
leading to the commutation relations
$$[\xmu ,p_\nu ] = -i{\delta^\mu}_\nu\;, \qquad
  \{d^A ,\thetaB \} ={\delta^A}_B\;, \qquad
  \{ \hatphiA ,\phi^B \} = {\delta_A}^B\;.\eqn\crels$$

The  wave-function is  taken to be a function of $\pmu$, $\thetaA$,
$\phiA$ together with an infinite set of ghost coordinates.
The spectrum is then found by seeking the BRST cohomology classes,
using the BRST operator that follows from the Batalin and Vilkovisky
procedure.

We now turn to the BRST cohomology of the model. The full analysis
requires the complete
 solution of the BV master equation with an
infinite number of ghost fields. However, as explained in [\mikea,\GH],
there is also a \lq small formalism' which only involves
the finite number of ghosts that are required for the symmetries of
the first kind (\ie\ those that act on the coordinates as well as the
gauge fields). It was shown in [\GH] that the BRST
cohomology of the full
formalism at zero ghost number is the same as that for the
BRST cohomology calculated with the small formalism, and for this reason
we shall restrict ourselves to physical states of zero
ghost number in this paper.
A state of  zero ghost number corresponds to a ghost
independent wave-function $\Phi =\Phi (p, \theta,\phi )$.  It was shown in
[\GH] that in this case $Q \Phi =0$ implies the constraints
$$\eqalign{& p^2 \Phi =0 \;, \qquad \psh d \Phi =0 \;, \qquad
			      \psh\hat\phi \Phi =0 \;,\cr
		 {\hat\phi}_A{\hat\phi}_B \Phi &=0 \;, \qquad
  [ d^A d^B -8(\hat\phi \Gammu )^{[A} (\Gamma_\mu \psh\phi )^{B]}]\Phi =0.
\cr}\eqn\condiux$$
Expanding
the wave-function in powers of
$\phi$,
$$\Phi (p,\theta ,\phi ) = \Psi_0 (p,\theta )
			  + \phiA\PsiA (p,\theta )
			   +\half\phiA\phiB\Psi_{AB} (p,\theta )
						+\dots ,\eqn\expanded$$
the constraint
${\hat\phi}_A{\hat\phi}_B \Phi =0$ implies that only $\Psi_0$ and
$\Psi_A$ are non vanishing. The constraint
$\Bigl[ d^A d^B
     -8(\hat\phi \Gammu )^{[A} (\Gamma_\mu \psh\phi )^{B]}\Bigr]\Phi =0$
implies that $\Psi_0$ satisfies $\dA \dB \Psi_0 =0$
and this implies that $\Psi_0$ is trivial (unless $\pmu =0$). Thus the
only non-trivial part of the wave-function is $\Psi_A$, and \condiux\ implies
that it satisfies precisely the covariant constraints \onenices,\twonices,
which lead to the Yang--Mills spectrum.
Note that the constraints \condiux\ are precisely what one expects
from taking the classical constraints \conditions, and requiring that
the corresponding operators annihilate the wave-function.

The $SO(8)$ light-cone gauge formalism of section 3 may be recovered
directly by fixing the gauge in the covariant action by setting $e=1$ and
the other gauge fields to zero and  imposing
$\xplus =\pplus\tau + \xplus_0 $, $\gamplus\theta =0$ and
$\gamplus \phi =0$.  The light-cone action
is given by $S_{lc}= S_{0lc} + S'_{lc}$, where
$$S_{0lc} =\inttau \Bigl[ p^i {\dot x}^i -\half p^i p^i
			+ i{\hat\theta}^{\adot}{\dot\theta}^{\adot}
			+ i{\hat\phi}^a {\dot\phi}^a \Bigr],\eqn\lcaction$$
and
$$S'_{lc} ={\half} \inttau\Bigl[\chi_{\adot\bdot}(d^{\adot} d^{\bdot}
	    + 2\pplus \gamma^{\adot\bdot}_{cd} \phi^c {\hat\phi}_d )
	    + \Upsilon^{ab}{\hat\phi}_a {\hat\phi}_b \Bigr],\eqn\conaction$$
where
$\chi_{\adot\bdot}$, $\Upsilon^{ab}$ are Lagrange multipliers
imposing the remaining constraints
$$d^{\adot} d^{\bdot}
      + 2\pplus \gamma^{\adot\bdot}_{cd} \phi^c {\hat\phi}_d =0\;, \qquad
			       {\hat\phi}_a {\hat\phi}_b =0,\eqn\subject$$
and are  gauge fields for a local $SO(8)\times SO(8)$ symmetry.

\section{Second type (linear constraint).}

In  this section,
a superparticle theory with a spinor wave-function satisfying the linear
constraint \spinlin\ will be considered.
We start with the superparticle action \pq -\tu\ formulated in an extended
superspace with coordinates $(x^\mu ,\theta_A ,\phi ^A )$ where $\theta_A$
and $\phi^A$ are anticommuting Majorana-Weyl spinors and add a lagrange
multiplier term imposing the constraint
${\dA} ({\Gamnrs} {)^B}_A {\hat\phi}_B =0$,
which then leads to the condition \spinlin\ on the wave-function. In fact,
to close the constraint algebra and obtain a gauge invariant superparticle
action, it is necessary to include a lagrange multiplier term
imposing the extra constraint $ \hat \phi  \hat \phi =0$.

The new superparticle action  is then given
by the sum of
$$S_0  =\inttau \Bigl[p_\mu \dot x^\mu +i\hat\theta\dot\theta
			+i\hat\phi \dot\phi \Bigr],\eqn\sspthree$$
and
$$\eqalign{S^{\prime\prime} =\inttau\Bigl[-\half ep^2  + i\psi\psh d
				   +i\varphi\psh\hatphi
				 & +i{\Lamnrs}\;d\;{\Gamnrs}\hat\phi \cr
		      &\qquad      -i\beta (\phi\hat\phi -1 )
				   +\half\hat\phi\omega\hat\phi \Bigr],
\cr}\eqn\xtraspin$$
where,
as usual, $p_\mu$ is the momentum conjugate to the space-time coordinate
$x^\mu$, $d^A$ is a spinor introduced so that the Grassmann coordinate
$\theta$ has a conjugate momentum ${\hat\theta}^A =d^A -\psh^{AB}\theta_B$,
$\phi^A$ is   a new spinor coordinate and $\hat\phi_A$ is its conjugate
momentum. The fields $e$, $\psi^A$, $\varphi_A$, $\Lamnrs$, $\beta$ and
${\omega}^{AB}$ are all Lagrange multipliers (which
are gauge fields for  corresponding local  symmetries) imposing the following
constraints
$$\eqalign{&p^2 =0, \qquad \psh d=0, \qquad \psh\hat\phi =0 \cr
	   {\hat\phi}_A{\hat\phi}_B =0,&\qquad
	    {\phiA}{\hat\phi}_A - 1 =0,\qquad
	    {\dA} ({\Gamnrs} {)^B}_A {\hat\phi}_B =0
\cr}\eqn\newthreecon$$

World-line reparameterization, when
combined with a  trivial symmetry, gives the {\cal A}  symmetry of \asymm.
There are two fermionic symmetries of the first kind, $\cal B$ and $\cal B'$
with fermionic spinor
parameters $\kappa^A (\tau)$ and $\zeta_A (\tau )$ given
by
$$\eqalign{\delta \psiA  &=\dot\kapa ,\qquad
	    \delta \theta  ={\psh}_{AB}\kappa^B ,\qquad
	    \delta e =4i{\dot\theta}_A \kapa ,\cr
	   &\delta x^\mu =i\dA ({\Gammu})_{AB}\kappa^B
		      +i\thetaA (\Gammu )^{AC}{\psh}_{CB}\kappa^B ,
\cr}\eqn\siegspin$$
and
$$\delta {\varphi}_A ={\dot\zeta}_A +\beta\zeta_A ,\qquad
  \delta \phiA =\zeta_B \psh^{BA} ,\qquad
  \delta x^\mu =i\hat\phi_A (\Gammu )^{AB}\zeta_B\eqn\ferspin$$
where
$\zeta_A$ is a spinor parameter. The
bosonic symmetries associated with the gauge fields $\beta$ and
${\omega}^{AB}$ (the {\cal C} and {\cal C}' symmetries) are defined
by
$$\eqalign{&\delta \beta =\dot\eta ,\qquad
	    \delta {\hat\phi}_A =\eta{\hat\phi}_A \;,\qquad
	    \delta \phiA = -\eta\phiA ,\cr
	    \delta \omega^{AB} =-2 & \eta\omega^{AB} ,\qquad
	    \delta \Lamnrs =\eta\Lamnrs \;,\qquad
	    \delta {\varphi}_A =-\eta{\varphi}_A \;,\cr}\eqn\etaspin$$
and
$$\delta \omega^{AB} ={\dot\Upsilon}^{AB} + 2\beta\Upsilon^{AB} ,\qquad
  \delta \phiA = i \Upsilon^{AB} {\hat\phi}_B ,\eqn\Upsispin$$
where
$\eta$ is a bosonic parameter and $\Upsilon^{AB} =-\Upsilon^{BA}$ is a
bosonic bispinor parameter. There is also a tensor symmetry associated
with the gauge field $\Lamnrs$ ({\cal F} symmetry) with
bosonic parameter $\Sigmnrs$ and given
by
$$\eqalign{&\delta \Lamnrs =\dot\Sigmnrs +\beta\Sigmnrs ,\qquad
	    \delta \dA = -2{\hat\phi}_B {\Sigmash^B}_C\psh^{CA} ,\cr
	   &\delta \thetaA = -{\hat\phi}_B {\Sigmash^B}_A \;,\qquad
	    \delta \phiA = d^B {\Sigmash_B}^A ,\qquad
	    \delta e = 4i{\hat\phi}_B {\Sigmash^B}_A\psiA ,\cr
    &\delta x^\mu = i{\hat\phi}_B {\Sigmash^B}_A (\Gammu )^{AC}\thetaC \;,
      \qquad \delta \omega^{AB} = 4i{\Sigmash^A}_C\psh^{CD}{\lmdash_D}^B .
\cr}\eqn\bosten$$

The covariant  quantization of this  model (in the gauge $e=1$ with other
gauge fields vanishing)   gives the classical    gauge-fixed action $$
  S_{fixed} =\inttau\Bigl[ p_\mu {\dot x}^\mu
		   -\half p^2 +i{\hat\theta}\dot\theta
			  +i\hat\phi \dot\phi\Bigr].\eqn\threefixed$$
Once again,  a complete treatment of the infinite number of ghosts requires
the solution of the BV master equation, which is not undertaken here and we
shall only consider  the ghost number zero cohomology class.
This is given by the wavefunction $\Phi =\Phi (x, \theta,\phi )$, and
$Q\Phi =0$ implies
that
$$\eqalign{& p^2 \Phi =0, \qquad
	      \psh d \Phi =0, \qquad
	       \psh\phi\Phi =0, \cr
       {\hat\phi}_A{\hat\phi}_B\Phi =&0, \qquad
	     (\phiA{\hat\phi}_A -1)\Phi =0,\qquad
	      \dA (\Gamnrs {)^B}_A{\hat\phi}_B \Phi =0.\cr}
\eqn\waveconstr$$
Expanding
the wavefunction in powers of $\phi$
$$
  \Phi (x,\theta ,\phi )=\Psi_0 (x,\theta )
			  +\phiA\PsiA (x,\theta )
			   +\half\phiA\phiB\Psi_{AB} (x,\theta ) +\dots \;,
\eqn\wavexpan$$
the
constraint $\hat\phi\hat\phi \Phi =0$ implies that only $\Psi_0$ and
$\PsiA$ are non vanishing, while $(\phi\hat\phi -1 )\Phi =0$ implies that
$\Psi_0$ is trivial. Hence, the only non-trivial part of the wavefunction
is $\PsiA$, and \waveconstr\ implies that it satisfies precisely the
covariant constraints \onenice\ and \twonice, which lead to the
super Yang--Mills spectrum.

Again the light-cone gauge equations of section 3 can be obtained directly
from the light-cone gauge action $S_{lc} = S_{0lc} + S_{lc}'$, where
$$S_{0lc} =\inttau\Bigl[ p^i {\dot x}^i -\half p^i p^i
		       +i{\hat\theta}^{\adot} {\dot\theta}^{\adot}
		       +i{\hat\phi}^a {\dot\phi}^a \Bigr],\eqn\lcnewnun$$
and
$$S'_{lc} =\inttau\Bigl[\Upsilon^{ab}{\hat\phi}_a {\hat\phi}_b
			  +\beta (\phi^a {\hat\phi}_a -1)
	       +{\Lamijk} d^{\adot} (\gammaijk {)^b}_{\adot} {\hat\phi}_b \;,
\Bigr].\eqn\xtranewnun$$
The Lagrange multipliers $\Upsilon^{ab}$,  $\beta$ and $\Lamijk$ impose
the required constraints.

\section {First ilk (extra vector coordinates).}

The
action \twisusy\
with world-line supersymmetry leads to a wave-function
$\Phi( x,\theta , \lambda_\alpha)$ and expanding this in $\lambda_\alpha$
gives $2^{[D/2]}$ superfields  which can be assembled into a spinor
wave-function $\Psi(x, \theta)_A$ ($A=1, \dots,2^{[D/2]}$)
satisfying the constraints \coco. The addition of the Lagrange multiplier
term,
$$S={1 \over 720} \inttau \Upsilon ^{ \mu \nu \rho}
		   \left(d \Gamma_{ \mu \nu \rho} d
		     + 4p_{[ \mu }\lambda_\nu \lambda_{ \rho]}\right),
\eqn\ilky$$
leads to the  constraint
$d \Gamma_{ \mu \nu \rho} d + 4p_{[ \mu }\lambda_\nu \lambda_{ \rho]}=0$.
Imposing this on the wave function gives  \spinquad\  when rewritten in
terms of $\Psi_A$  so that the spectrum of Yang--Mills is again obtained.
This is the Lagrangian formulation of the first ilk superparticle of
[\rocek ]. The full Batalin-Vilkovisky quantization of this model was
given in [\spaces,\spacesa], confirming that
 its spectrum  is indeed that of the
Yang--Mills supermultiplet.

\chapter{Superparticles with vector super wave-functions.}

In
this section, we shall consider superparticle theories which are formulated
in an extended superspace with coordinates $(x^\mu ,\theta_A ,\phimu )$
where $\theta_A$ is an anticommuting Majorana-Weyl spinor, and $\phimu$
is a new commuting vector coordinate. The model  defined by \pq, \vw\ was
shown to lead
to a vector wave-function   satisfying the constraints \veccon\ and
identified modulo \modgag. We now wish to modify this to obtain models
with wave-functions satisfying the extra constraints \quadratic\ or \linear.
These will be referred to as models of the third and fourth type,
respectively.

\section{Third type (quadratic constraint)}

As
in the first type of model discussed in the last section, there are two
versions of this model, one with the constraint $N=1$ and one with the
constraint $\hat \phi  \hat \phi =0$. Here we shall just consider the
latter version, which leads to a vector wavefunction satisfying the
quadratic constraint \quadratic .
The  new superparticle action is given
by
$$S_{0} =\inttau \Bigl[p_\mu {\dot x}^\mu + i\hat\theta\dot\theta
		       +{\hatphi}_\mu {\dot\phi}^\mu \Bigr],\eqn\newvecci$$
plus the term
$$\eqalign{S_{1}=\inttau \Bigl[-\half ep^2 + i\psi\psh d + \half d\chi d
			     & +2i{\hatphi}_\mu\chi{{\psh}^\mu}_\nu\phinu\cr
			     & -\omega \pmu {\hatphi}_\mu
			       +\half {\hatphi}_\mu\;\rhomunu{\hatphi}_\nu
			      \Bigr],
\cr}\eqn\xtravecci$$
where
we use the convenient
notation
$${\psh}_{\mu \nu}^{AB}= p^\rho \Gamma _{\mu \rho \nu}^{AB} .\eqn\connot$$
The
fields $e$, $\psi^A$, $\chi_{AB} =-\chi_{BA}\;$,
$\rho_{\mu\nu} =\rho_{\nu\mu}$ and $\omega$ are all Lagrange multipliers
imposing the following
contraints
$$\eqalign{&p^2 =0, \qquad \psh d=0, \qquad p^\mu {\hatphi}_\mu =0, \cr
	    {\hatphi}_\mu {\hatphi}_\nu &=0, \qquad
	    \dA \dB + 4 {\hatphi}_\mu ({\psh^\mu}_\nu )^{AB} \phinu =0,
\cr}\eqn\verdevec$$
and
are also gauge fields for their corresponding symmetries.

The action \newvecci -\xtravecci\ is invariant under
the global space-time supersymmetry
transformations
$$\delta\theta =\epsilon ,\qquad
  \delta x^\mu =i\epsilon\Gamma^\mu\theta ,\eqn\globalvec$$
where
$\epsilon$ is a Grassmann parameter, together with a number of
local symmetries of the first-kind. World-line reparameterization, when
combined with a   trivial symmetry gives the {\cal A}
transformations, \asymm.  The fermionic symmetry, $\cal B$, with fermionic
spinor parameter
$\kappa^A (\tau)$, is given
by
$$\eqalign{\delta \psi^A &=\dot\kapa ,\qquad
	   \delta \theta_A =\psh_{AB}\kappa^B ,\qquad
	   \delta e =4i{\dot\theta}_A \kapa ,\cr
	  &\delta x^\mu =i\dA ({\Gammu})_{AB}\kappa^B
		      +i\theta_A(\Gammu )^{AC}\psh_{CB}\kappa^B .
\cr}\eqn\fermvec$$
and the
bosonic symmetries associated with the gauge fields $\omega$ and
$\chi$ (the {\cal C} and {\cal C}'  symmetries, respectively)
are defined
by
$$\delta \omega =\dot\zeta ,\qquad
  \delta \phimu =\zeta p^\mu ,\qquad
  \delta x^\mu =\zeta \hatphi^\mu ,\qquad
  \delta e =4i\zeta (\Gammu )^{AB}\chi_{AB} \hatphi_\mu ,\eqn\bosveczet$$
and
$$\eqalign{&\delta \dA =2i\psh^{AB}\lambda_{BC}\;\dC ,\qquad
	 \delta{\rhomunu}
		= -4i({\psh_{(\mu}}^\sigma )^{AB}\lambda_{AB}\;
		     \rhosignu\;,\cr
	&\delta \phimu =-2i\lambda_{AB}({\psh^\mu}_\nu)^{AB} \phinu \;,\qquad
	 \delta {\hatphi}_\mu
		= 2i{\hatphi}_\nu\lambda_{AB}({\psh^\nu}_\mu )^{AB}\;,\cr
	&\delta \thetaA = -i\dB\lambda_{BA}\;,\qquad
	 \delta \chi_{AB} ={\dot\lambda}_{AB}
			+4i\lambda_{AC}\psh^{CD}\chi_{DB} ,\cr
	&\delta x^\mu =-\dA\lambda_{AB}(\Gammu )^{BC}\thetaC
		 -2i{\hatphi}_\sigma\lambda_{AB}(\Gamsmn )^{AB} \phinu\;,\cr
	&\delta e =-4\dA\lambda_{AB}\psi^B
		   -4i{\hatphi}_\mu (\Gammu )^{AB}\lambda_{AB}\;\omega\;,
\cr}\eqn\bosveclam$$
where
$\lambda_{AB} =-\lambda_{BA}$ is a bosonic bispinor parameter.
There is also a tensor symmetry associated with the gauge field
$\rho_{\mu\nu}$ (the {\cal E} symmetry) and given
by
$$\delta \rhomunu ={\dot u}^{(\mu\nu )}
		   +4iu^{(\mu\sigma}\chi_{AB}({\psh_\sigma}^{\nu )})^{AB}
		      \;,\qquad
  \delta \phimu = - u^{(\mu\nu )}\hatphi_\nu ,\eqn\bosten$$
where
$u_{\mu\nu}=u_{\nu\mu}$ is a tensor parameter.

The classical  gauge-fixed action (in the gauge with  $e=1$ and other gauge
fields vanishing)  here takes the form
$$S =\inttau \Bigl[ p_\mu{\dot x}^\mu -\half p^2
		  + i{\hat\theta}{\dot\theta} + \hatphi_\mu \dot\phi^\mu
\Bigr].\eqn\verdefixed$$

The wave-function $\Phi =\Phi (x ,\thetaA ,\phimu )$ is a function
of $\xmu$, $\thetaA$, and $\phimu$ (again we are considering only the
zero-ghost sector).  The constraint $Q\Phi =0$ implies
$$\eqalign{&p^2 \Phi =0, \qquad \psh d \Phi =0, \qquad
	    p_\mu \phimu \Phi =0, \cr
	   {\hatphi}_\mu &{\hat\phi}_\nu \Phi =0, \qquad
	   [\dA \dB + 4{\hatphi}_\mu ({\psh^\mu}_\nu )^{AB} \phinu ]\Phi=0.
\cr}\eqn\waverde$$
Expanding
the wavefunction in powers of $\phi$
$$\Phi (x,\thetaA ,\phimu ) = \Psi_0 (x,\thetaA )
			     +\phimu \Psimu (x,\thetaA )
			     +\half \phimu\phinu \Psi_{\mu\nu} (x,\thetaA )
			     +\dots ,\eqn\powerverde$$
the constraint
${\hatphi}_\mu {\hatphi}_\nu \Phi=0$  implies that only $\Psi_0$ and
$\Psimu$ are non vanishing. The condition
$(d d +4{\hatphi}_\mu {\psh^\mu}_\nu\phinu )\Phi=0$ implies
$d d \Psi_0 =0$, and hence that $\Psi_0$ is trivial
(if $\pmu \not=$0).
Thus, the only non-vanishing part of the wave-function is $\Psimu$, and
\waverde\ implies that it satisfies precisely the covariant constraints
\onequadr\ and \twoquadr\ which lead to the physical spectrum of the
super Yang--Mills theory.

The light-cone action is  $S_{0lc}+ S'_{lc}$ with
$$S_{0lc} =\inttau \Bigl[ p^i {\dot x}^i -\half p^i p^i
		       + i{\hat\theta}^{\adot} {\dot\theta}^{\adot}
		       + \hatphi^i {\dot\phi}^i
\Bigr],\eqn\lcverde$$
and
$$S'_{lc} ={\half} \inttau \Bigl[ \rhoij \hatphi_i \hatphi_j
	     + \chi_{\adot\bdot} \Bigl( d^{\adot} d^{\bdot}
	     - 4\pplus {\hatphi}_i ({\gammai}_j )^{\adot\bdot}\phi^j \Bigr)
\Bigr],\eqn\verdeconst$$
where
$\rho_{ij}$ and $\chi_{\adot\bdot}$
are Lagrange multipliers that impose the quadratic constraints.
Applying these constraints to the light-cone wave function leads directly
to the conditions \quadratic.

\section {Fourth type (linear constraint)}

A  superparticle theory with  a vector wave-function satisfying the
linear constraint \linear \
is obtained from the action \pq, \vw\ formulated in the superspace with
coordinates $(\xmu,\thetaA ,\phimu )$ together with a Lagrange multiplier
term that imposes the
 constraint $\Cmupha =0$, where
$$\Cmupha = \hatphi_\mu \dA
	   - \seventh \hatphi_\nu ({ {\Gamma_\mu}^\nu }{)^A}_B \dB
.\eqn\eiey$$
The complete superparticle action is then given by  $S_0 + S'$ where,
$$S_0 =\inttau \Bigl[ p_\mu {\dot x}^\mu
			      +i{\hat\theta}{\dot\theta}
			      +\hatphi_\mu {\dot\phi}^\mu
\Bigr],\eqn\azulzero$$
and
$$\eqalign{S' =\inttau \Bigl[-\half e p^2  + & i\phi\psh d
			      +\lambda \pmu {\hat\phi}_\mu
			   +\half \hatphi\omega\hatphi \cr
			    & +\beta (\phimu \hat\phi_\mu -1)
			      +\Upsimupha \Cmupha \Bigr].
\cr}\eqn\azulxtra$$

The action is again invariant under ${\cal A}$ transformations as well as
fermionic symmetries associated with the gauge fields $\psiA$,
$\Upsimupha$ (the {\cal B} and {\cal B}' symmetries respectively )
which are defined
by
$$\eqalign{\delta \psiA &= {\dot\kappa}^A\;, \qquad
	   \delta \thetaA =\psh_{AB}\kappa^B\;,\qquad
	   \delta e = 4i{\dot\theta}_A\kappa^A         \cr
	  &\delta \xmu = i\dA (\Gammu )_{AB}\kappa^B
			+i\thetaA (\Gammu )^{AC}\psh_{CB}\kappa^B\;.
\cr}\eqn\kissy$$
and
$$\eqalign{&\delta\Upsimupha ={\dot\chi^\mu}_A +\beta\chimuA\;,\cr
	   &\delta\phimu =-\chimuA\dA
			  + {\seventh}\chinuA{\Gamma_\nu}^\mu\dA\;  \cr
	   &\delta\thetaA =-i\chimuA\hatphi_\mu
			  +{\seventh}i\chimuA\hatphi_\nu{\Gamma_\mu}^\nu\;,\cr
	   &\delta\xmu =-[\chinuA\hatphi_\nu
			  -{\seventh}\chisigA\hatphi_\nu{\Gamma_\sigma}^\nu ]
			   (\Gammu\theta )^A\;,\cr
	   &\delta\dA =-2i[\chimuB\hatphi_\mu
			   -{\seventh}\chimuB\hatphi_\nu{\Gamma_\mu}^\nu ]
			    {\psh}^{BA}\;,\cr
	   &\delta e= -4[\chimuA\hatphi_\mu
			 -{\seventh}\chimuA\hatphi_\nu{\Gamma_\mu}^\nu ]
			 \psiA\;.
\cr}\eqn\upsikiss$$
The bosonic symmetries associated with the gauge fields $\lambda$ and
$\beta$ ({\cal C} and {\cal C}' symmetries, respectively ) are defined
by
$$\delta\lambda =\dot\zeta +\beta\zeta\;, \qquad
  \delta\xmu =-\zeta\hatphi^\mu\;, \qquad
  \delta\phimu = -\zeta\pmu\;,           \eqn\kissyC$$
and
$$\eqalign{\delta\beta =&\dot\eta\;,\qquad
	   \delta\hatphi_\mu =\eta\hatphi\;,\qquad
	   \delta\phimu =-\eta\phimu\;,    \cr
	  &\delta\lambda =-\eta\lambda\;,\qquad
	   \delta\omegmunu =-2\eta\omegmunu\;,
\cr}\eqn\kissyD$$
There is also a tensor
symmetry associated with the gauge field $\omegmunu$ (the {\cal F} symmetry)
with bosonic parameter $\Sigmunu$ and defined
by
$$\delta\omegmunu =\dot\Sigmunu + 2\beta\Sigmunu\;,\qquad
  \delta\phimu = - \Sigmunu\hatphi_\nu\;.\eqn\kissyF$$

The classical auge-fixed action (in the  $e=1$ gauge with the other gauge
fields set to zero) takes the form
$$S_{fixed} =\inttau \Bigl[ p_\mu {\dot x}^\mu -\half p^2
			   + i\hat\theta\dot\theta
			   + \hatphi_\mu \dot\phi^\mu
\Bigr].\eqn\azulfixed$$
The wave-function is a function of $\pmu$, $\thetaA$
and $\phimu$ together with the infinite set of ghost coordinates.
Once again the zero ghost-number wavefunction $\Phi =\Phi (p,\theta ,\phi )$
satisfying   $Q\Phi =0$
satisfies constraints that arise from the variation of the Lagrange
multipliers in the action.
$$
  \eqalign{&p^2 \Phi =0 ,\qquad \psh d\;\Phi =0 ,\qquad
	    \pmu\hatphi_\mu\;\Phi =0, \cr
	    \hatphi_\mu \hatphi_\nu \;&\Phi=0,\qquad
	    {C_\mu}^A\;\Phi =0,\qquad (\phi^\mu \hatphi_\mu -1)\Phi =0
\cr}.\eqn\azulcohom$$
Expanding
the wavefunction in powers of $\phi$
$$\Phi (p,\theta ,\phi ) = \Psi_0 (p,\theta )
			  +\phimu\Psi_\mu (p,\theta )
			  +\half \phimu\phinu \Psi_{\mu\nu} (p,\theta )
			  +\dots ,\eqn\azulpower$$
the
constraint $\hatphi_\mu\hat\phi_\nu \Phi =0$ implies that only
$\Psi_0$ and $\Psi_\mu$ are non-vanishing. The constraint
$(\phi\hatphi -1)\Phi =0$  makes  $\Psi_0$ trivial. Hence, the
only non-trivial part of the wavefunction is $\Phi_\mu$, and \azulcohom\
implies that it satisfies precisely the covariant constraints
\onelinear -\twolinear . We have thus shown that the BRST cohomology
class with no ghost dependence gives the physical spectrum of the
super Yang--Mills theory.

The  light-cone action (in which $x^{+} =\pplus\tau + {x_0}^{+}$,
$\gamplus\theta =0$ and $\phi^{+} =0$) is given by $S_{0lc} + S'_{lc}$ with
$$S_{lc} =\inttau \Bigl[ p^i {\dot x}^i -\half p^i p^i
			+i{\hat\theta}^{\adot}{\dot\theta}^{\adot}
			+\hatphi^i {\dot\phi}^i \Bigr],\eqn\azulcone$$
and
$$S'_{lc} =\half \inttau \Bigl[ \omega^{ij} \hatphi_i \hatphi_j
				 + \beta (\phi^i \hatphi^i -1)
				 + \Upsilon^{i\adot} C^{i\adot}
\Bigr],\eqn\azullybit$$
where
$\omega^{ij}$, $\beta$ and $\Upsilon^{i\adot}$ are Lagrange
multipliers imposing constraints that lead to the linear condition \linear\
on the light-cone gauge vector  wave function.

\

\ack

{The early parts of this work were done in collaboration with Michael Green
and we would like to thank him for many helpful discussions
and comments.}

\refout

\end
\bye